\documentclass{elsart}
\usepackage{amssymb,epsf}

\newcommand{\Tr}{{\bf Tr}}

\newcommand{\contra}[2]{\begin{picture}(36,10)
  \put(0,0){$#1$}
  \put(2,-5){\line(0,1){2}}
  \put(2,-5){\line(1,0){34}}
  \put(36,-5){\line(0,1){2}}
  \put(34,0){$#2$} 
  \put(36,-7){\tiny $\beta$}
\end{picture}}

\newcommand{\contrad}[2]{\begin{picture}(40,10)
  \put(0,0){$#1$}
  \put(2,-5){\line(0,1){2}}
  \put(2,-5){\line(1,0){38}}
  \put(40,-5){\line(0,1){2}}
  \put(38,0){$#2$} 
  \put(40,-7){\tiny $\beta$}
\end{picture}}

\newcommand{\contradd}[2]{\begin{picture}(54,10)
  \put(0,0){$#1$}
  \put(2,-5){\line(0,1){2}}
  \put(2,-5){\line(1,0){52}}
  \put(54,-5){\line(0,1){2}}
  \put(52,0){$#2$} 
  \put(54,-7){\tiny $\beta$}
\end{picture}}

\newcommand{\contraddd}[2]{\begin{picture}(66,10)
  \put(0,0){$#1$}
  \put(2,-5){\line(0,1){2}}
  \put(2,-5){\line(1,0){64}}
  \put(66,-5){\line(0,1){2}}
  \put(64,0){$#2$} 
  \put(66,-7){\tiny $\beta$}
\end{picture}}

\begin{document}

\begin{frontmatter}
\title{Enhancement of Critical Slowing Down 
in Chiral Phase Transition \\
--- Langevin Dynamics Approach ---}

\author{T.~Koide}
\address{Institute f\"ur Theoretische Physik, J.~W.~Goethe Universit\"at, 
D-60054 Frankfurt, Germany}
\author{M.~Maruyama}
\address{Department of Physics, Tohoku University, Sendai 980-8578, Japan}

\begin{abstract}
We derive the linear Langevin equation 
that describes the behavior of the fluctuations of the order parameter of 
the chiral phase transition above the 
critical temperature by applying the projection operator method to 
the Nambu-Jona-Lasinio model at finite temperature and density.
The Langevin equation relaxes exhibiting oscillation, 
reveals thermalization and 
converges to the equilibrium state consistent with the mean-field approximation 
as time goes on.
With the help of this Langevin equation, 
we further investigate the relaxation of the critical fluctuations.
The relaxation time of the critical fluctuations 
increases at speed as the temperature approaches toward 
the critical temperature because of the critical slowing down.
The critical slowing down is enhanced as the chemical potential 
increases because of the Pauli blocking.
Furthermore, we find another enhancement of the critical slowing down 
around the tricritical point.
\end{abstract}


\end{frontmatter}

\section{Introduction}

It is widely believed that 
hadronic matters undergo a phase transition at high temperature, 
and hence become the quark-gluon plasma (QGP), where 
quarks and gluons are deconfined and chiral symmetry broken in the hadron phase 
is restored.
Relativistic heavy-ion collisions provide significant opportunities 
to explore the QGP.
Heavy-ion collisions are essentially nonequilibrium processes and hence 
the description of the time evolution is necessary to understand 
the phenomena in a comprehensive way.
Then, the hydrodynamic model is a powerful tool, 
and fairly correctly describes 
the experimental data of heavy-ion collisions, for example, 
transverse momentum dependence of the elliptic flow coefficient $v_{2}$
\cite{ref:Bearden,ref:Roland,ref:Kolb}.

Recently, our attention has been focused 
on dynamics near the critical points of the chiral phase transition
\cite{ref:Raja,ref:Raja2,ref:SS1,ref:SS2,ref:Rischke,ref:Hatta,ref:Fujii,ref:Fujii2,ref:Paech,ref:Kodama,ref:Boya0,ref:Boya1,ref:Boya2,ref:Boya3,ref:Boya}.
Unfortunately, the hydrodynamic model (in particular, the so-called ideal fluid) 
will be inadequate to describe 
nonequilibrium phenomena near the critical point.
Because it is valid only in the case where 
we are interested in gross variables 
associated with macroscopic time and length scales 
(entropy density, energy density, etc) 
and 
they are widely separated from other variables 
associated with microscopic scales(degrees of freedom of each quark and gluon).
To understand it, we should remember the dynamical hierarchy 
of classical dilute gas that is shown in TABLE \ref{tab:1}
\cite{ref:Kawasaki,ref:Kawasaki2,ref:Kawasaki3}.
The classical gas consists of a lot of classical particles and 
the dynamics is described by the Liouville equation.
Then, the typical length and time scales are given by $r_0$ and $r_0 / v_{th}$, 
respectively, where $r_0$ is the interaction length and $v_{th}$ is 
the thermal averaged velocity.
However, when we are interested in slow motions associated 
with the scales of the mean free path $L$ and $L/v_{th}$, 
the Liouville equation is reduced 
to the kinetic equation like the Boltzmann equation.
Furthermore, 
when we observe quantities associated with further slower scales 
of the wave length and frequency of sound $L_{macro}$ and $T_{macro}$, 
the Boltzmann equation is reduced to the hydrodynamic equation.
Thus, in order to apply the hydrodynamic model, 
the microscopic scale must be smaller than the macroscopic scale, 
$\epsilon = r_0 / L_{macro} << 1$.
As a matter of fact, we assume the local equilibrium in the hydrodynamic equation 
and the deviation from the local equilibrium can be characterised by $\epsilon$.
For example, the number density $\langle n({\bf x},t) \rangle_{ini}$ 
is approximated as 
\begin{eqnarray*}
\langle n({\bf x},t) \rangle_{ini} 
= \langle n({\bf x}) \rangle_{local} + O(\epsilon),
\end{eqnarray*}
where $\langle~~\rangle_{ini}$ means an expectation value by an initial state and 
$\langle~~\rangle_{local}$ denotes that by a local equilibrium state.
When we completely ignore the deviation from the local equilibrium, 
we obtain the ideal fluid.
When we take into account the deviation from the local equilibrium, 
we obtain the dissipation term and the Euler equation is replaced by 
the Navier-Stokes equation
\cite{ref:Muronga}.

Now, we return to the discussion of the critical dynamics.
In the following, we assume that there exists 
a similar dynamical hierarchy 
even in quantum field theory because of the success of the hydrodynamic model 
in relativistic heavy ion collisions, although we do not have clear evidence so far
\cite{ref:Cooper}.
On the other hand, 
to apply the hydrodynamic model for describing the critical dynamics, 
there must exist the clear separation between the microscopic scale and 
the macroscopic scale, as is discussed above.
However, near the critical points, the microscopic correlation length 
indefinitely increases and hence the expansion parameter $\epsilon$ is not 
small any more.
Therefore, we cannot apply the hydrodynamic model to describe 
the critical dynamics.
On the other hand, it is probable that 
we still do not need the full details of the 
Heisenberg equation of motion because of the universality of phase transition.
Then, from an analogy to gas dynamics, 
the critical dynamics can be regarded as a sort of mesoscopic scale dynamics 
and will be described by a kinetic equation.
As a matter of fact, it is known that 
the critical dynamics in condensed matter physics 
is described by the Langevin equation.
This theory is called the mode coupling theory
\cite{ref:Kawasaki,ref:Kawasaki2,ref:Kawasaki3,ref:HH,ref:Onuki}.

\begin{table}\leavevmode
\begin{center}
\begin{tabular}{cccc}
Region & Equation & length scale & time scale \\ \hline 
\hline
microscopic  & Liouville & $r_0$ & $r_0/v_{th}$ \\
mesoscopic   & Boltzmann & $L$ & $L/v_{th}$ \\ 
macroscopic  & Hydrodynamic & $L_{macro}$ & $T_{macro}$  \\ \hline
\hline
\end{tabular}
\caption{The dynamical hierarchy of classical dilute gas.
$r_0$ is the interaction length, $v_{th}$ is the averaged 
thermal velocity and $L$ is the mean fee path. 
$L_{macro}$ and $T_{macro}$ mean macroscopic scales like 
wave length and frequency of sound.}
\label{tab:1}
\end{center}
\end{table}

One of the aims of this paper is to derive a linear Langevin equation 
for describing the dynamics of the critical fluctuations above the 
critical temperature of the chiral phase transition.
There are two typical methods to derive the Langevin equation from a 
microscopic point of view
\cite{ref:Wakou}.
One is the variational method that is applied to a sort of 
an effective action 
\cite{ref:Morikawa,ref:Glei-Ramo,ref:Ber-Glei-Ramo,ref:Lom-Mazz,ref:Grei-Mul,ref:Grei-Leu,ref:Xu-Grei,ref:Risch}.
Then, the noise is introduced as an auxiliary field. 
However, as is discussed in 
\cite{ref:KMT2}, there is arbitrariness for 
the introduction of the noise in this approach.

In this paper, we use the projection operator method.
This method was proposed by Nakajima
\cite{ref:Naka}, Zwanzig
\cite{ref:Zwanzig1} 
and Mori
\cite{ref:Mori}, and developed from various points of view 
\cite{ref:KMT2,ref:SH1,ref:SH2,ref:SH3,ref:SH4,ref:KMT1,ref:KM1,ref:Koide,ref:KM2,ref:Rau1,ref:Rau2,ref:Rau3,ref:Ayik,ref:Ana,ref:Howard,ref:review,ref:review2}.
In the following, we will discuss based on the formulation evolved by 
Shibata, Hashitsume, Uchiyama and the authors
\cite{ref:KM1}.
The derivation of the Langevin equation at vanishing chemical potential 
was discussed in 
\cite{ref:KM2}.
In this paper, we extend the result to derive the Langevin equation 
at finite temperature and density.
We show that our Langevin equation reveals thermalization and 
converges to the equilibrium state consistent with the mean-field approximation 
as time goes on.

The other aim is to investigate the critical slowing down 
in the chiral phase transition
\cite{ref:vanHove}.
As is well-known, the long wave length component of the order parameter 
shows extraordinary large fluctuations near the critical point 
and the increase of the relaxation time.
This slowing down of the relaxation of 
the critical fluctuations is known as the critical slowing down.
We define the relaxation time of the critical fluctuations and 
investigate the temperature and chemical potential dependences.
Then, we find the enhancement of the critical slowing down 
in the low temperature and large chemical potential region and 
around the tricritical point.

This paper is organized as follows:
In Sec. \ref{chap:POM}, we review the projection operator method 
following 
\cite{ref:KM1}.
In Sec. \ref{chap:Mori}, we introduce the Mori projection operator.
In Sec. \ref{chap:NJL}, we apply the projection operator method to the 
Nambu-Jona-Lasinio (NJL) model
\cite{ref:HK-PR,ref:HK-PR2}.
The equation obtained in Sec. \ref{chap:NJL} is not yet the Langevin equation.
To derive the Langevin equation, we employ the renormalization of the memory 
function in Sec. \ref{chap:renor}.
In Sec. \ref{chap:PS}, we calculate the power spectrum and 
discuss that our Langevin equation is consistent with 
the well-known results in equilibrium: the critical temperature, thermalization 
and the soft mode.
In Sec. \ref{chap:CSD}, 
we discuss the critical slowing down near the critical points 
by using the Langevin equation, and show that there exists the enhancement of 
the critical slowing down in low temperature and large chemical potential region 
and around the tricritical point.
Concluding remarks are given in Sec. \ref{chap:Sum}.

\section{Projection operator method} \label{chap:POM}

In this section, we review the projection operator method.
There are two different approaches to derive coarse-grained equations 
in the projection operator method.
When we apply this method in the Heisenberg picture, 
we can derive the Langevin equation.
On the other hand, 
the master equation is derived in the Schr\"odinger picture
\cite{ref:SH4}.
In this paper, we discuss in the Heisenberg picture, 
following 
\cite{ref:KM1}.
In the quantum field theory, 
the time evolution of operators is governed by the Heisenberg equation of motion,
\begin{eqnarray}
  \frac{d}{dt}O(t) &=& i[H,O(t)] \\
                   &=& iLO(t) \\
   \longrightarrow O(t) &=& e^{iL(t-t_{0})}O(t_{0}), \label{eqn:HE}
\end{eqnarray}
where $L$ is the Liouville operator and $t_{0}$ is an initial time at which 
we prepare an initial state.
The Heisenberg equation contains the information 
not only of gross variables but also of microscopic variables.
The latter is irrelevant information and we carry out coarse-grainings 
by introducing a projection operator $P$.
The projection operator $P$ and its complementary operator $Q = 1-P$ 
have the following general properties:
\begin{eqnarray}
   P^2 &=& P, \\
   PQ &=& QP = 0.
\end{eqnarray}
From Eq. (\ref{eqn:HE}), one can see that the time dependence of operators 
is determined by $e^{iL(t-t_{0})}$.
This operator obeys the differential equation,
\begin{eqnarray}
   \frac{d}{dt}e^{iL(t-t_{0})} 
   = e^{iL(t-t_{0})}iL 
   = e^{iL(t-t_{0})}(P+Q)iL.   \label{eqn:P+Q}
\end{eqnarray}
From this equation, we can derive the following two equations:
\begin{eqnarray}
  \frac{d}{dt}e^{iL(t-t_{0})}P 
&=& e^{iL(t-t_{0})}PiLP +e^{iL(t-t_{0})}QiLP, \label{eqn:P}\\
  \frac{d}{dt}e^{iL(t-t_{0})}Q 
&=& e^{iL(t-t_{0})}PiLQ +e^{iL(t-t_{0})}QiLQ. \label{eqn:Q}
\end{eqnarray}
Equation (\ref{eqn:Q}) can be solved for $e^{iL(t-t_{0})}Q$,
\begin{eqnarray}
e^{iL(t-t_{0})}Q 
= 
Qe^{iLQ(t-t_{0})} 
+ \int^{t}_{t_{0}}d\tau e^{iL(\tau-t_{0})}PiLQe^{iLQ(t-\tau)} 
\label{eqn:TC-dainyuu}.
\end{eqnarray}
Substituting Eq. (\ref{eqn:TC-dainyuu}) into Eq. (\ref{eqn:P+Q}) 
and operating $O(t_{0})$ from the right, 
we obtain the time-convolution (TC) equation,
\begin{eqnarray}
  \frac{d}{dt}O(t) 
&=& e^{iL(t-t_{0})}PiLO(t_{0}) +\int^{t}_{t_{0}}d\tau e^{iL(t-\tau)}
PiLQe^{iLQ(\tau-t_{0})}iLO(t_{0}) \nonumber \\
&&+ Qe^{iLQ(t-t_{0})}iLO(t_{0}). \label{eqn:TC-1}
\end{eqnarray}
The first term on the r.h.s. of the equation is called the streaming term 
and corresponds to a collective oscillation such as plasma wave, spin wave 
and so on.
The second term is the memory term that causes dissipation.
The third term is the noise term.
We can show that 
the memory term can be expressed as the time correlation of the noise.
This relation is called 
the fluctuation-dissipation theorem of second kind (2nd F-D theorem).
See Appendix \ref{app:FD} for details.
The elimination of microscopic variables using the projection operator 
gives rise to the dissipation term and the noise term in the coarse-grained 
macroscopic equation.

The TC equation is still equivalent to the Heisenberg equation and 
we cannot solve it exactly in general.
In order to approximate the memory term, we reexpress it as 
\begin{eqnarray}
\int^{t}_{t_{0}}d\tau 
e^{iL(t-\tau)}PiLQ{\mathcal D}(\tau,t_{0})e^{iQL_{0}Q(\tau-t_{0})}iLO(t_{0}),
\end{eqnarray}
where 
\begin{eqnarray}
{\mathcal D}(t,t_{0}) 
&=& 1+\sum^{\infty}_{n=1} i^n \int^{t}_{t_{0}}dt_{1}\int^{t_{1}}_{t_{0}}dt_{2} \cdots \int^{t_{n-1}}_{t_{0}}dt_{n} \nonumber \\
&&\hspace*{-1cm}\times Q\breve{L}^{Q}_{I}(t_{n}-t_{0})Q\breve{L}^{Q}_{I}(t_{n-1}-t_{0}) \cdots 
Q\breve{L}^{Q}_{I}(t_{1}-t_{0}), \nonumber \\
\end{eqnarray}
and
\begin{eqnarray}
\breve{L}^{Q}_{I}(t-t_{0}) 
&\equiv& e^{iQL_{0}Q(t-t_{0})}L_{I}e^{-iQL_{0}Q(t-t_{0})}.
\end{eqnarray}
Here, we introduced $L_{0}$ and $L_{I}$, that are Liouville operators of the 
nonperturbative Hamiltonian $H_{0}$ and 
the interaction Hamiltonian $H_{I}$, respectively,
\begin{eqnarray}
L_0~O = [H_0,O],~~~
L_{I}~O = [H_I,O].
\end{eqnarray}
When we expand ${\mathcal D}(t,t_0)$ up to first order in terms of $L_{I}$, 
we have
\begin{eqnarray}
\frac{d}{dt}O(t_{0}) 
&=& e^{iL(t-t_{0})}PiLO(t_{0}) + \int^{t}_{t_{0}}d\tau e^{iL(t-\tau)}
PiLQe^{iQL_{0}Q(\tau-t_{0})}iLO(t_{0}) 
+
\xi(t). \nonumber \\
\label{eqn:TC-2}
\end{eqnarray}
The noise term $\xi(t)$ should be determined so as to satisfy 
the 2nd F-D theorem, as we will see later.
Equation (\ref{eqn:TC-2}) is the starting point in the following calculation.

\section{Introduction of the Mori projection}\label{chap:Mori}

\begin{figure}\leavevmode
\begin{center}
\epsfxsize=7cm
\epsfbox{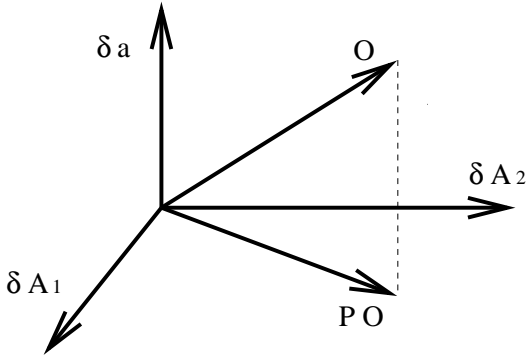}
\caption{The schematic figure of the Mori projection operator.
An arbitrary operator $O$ is projected onto the space 
spanned by gross variables $\delta A_{i}$.
The variable $\delta a$ means a microscopic variable.}
\label{fig:pro}
\end{center}
\end{figure}

In the derivation of the TC equation, we have not fixed a projection operator.
If we can solve the TC equation exactly, the final result does not depend on the 
projection operator because the TC equation 
is equivalent to the Heisenberg equation.
However, normally, it is impossible to solve the TC equation exactly, and 
we must start from the approximated equation (\ref{eqn:TC-2}).
Then, the choice of the projection operator is important.
There are several possible projection operators that extract slowly varying 
parts from an operator.
In this study, we adopt the Mori projection operator.
The Mori projection operator projects any operators onto the space 
spanned by gross variables.
There are three candidates for gross variables:
(i) order parameters, (ii) density variables of conserved quantities and 
(iii) their products.
When we can find out a complete set of gross variables, 
the macroscopic time evolution will be approximated by the superposition of the 
gross variables, as is shown in Fig. \ref{fig:pro}, schematically.
For this purpose, we introduce the Mori projection operator defined by 
\begin{eqnarray}
P~O = \sum_i c_i A_i,
\end{eqnarray}
where $A_i$ is the complete set of gross variables and 
the coefficient $c_i$ is given by 
\begin{eqnarray}
c_i = \sum_j (O,A_j) \cdot (A,A)^{-1}_{ji}.
\end{eqnarray}
The inner product is the Kubo's canonical correlation, 
\begin{eqnarray}
(X,Y) = \int^{\beta}_{0} \frac{d \lambda}{\beta}
{\rm Tr}[\rho~ e^{\lambda H}Xe^{-\lambda H}Y], \label{eqn:KCC}
\end{eqnarray}
where $\rho = e^{-\beta H}/{\rm Tr}[e^{-\beta H}]$ 
with the temperature $\beta^{-1}$. 
The inverse of the canonical correlation is defined by 
\begin{eqnarray}
\sum_j (A,A)^{-1}_{ij} \cdot (A_j,A_k)
= \delta_{i,k}.
\end{eqnarray}
The physical meaning of the Mori projection operator is discussed in Appendix 
\ref{app:MPRP} in detail.

\section{Linear Langevin equation in the Nambu-Jona-Lasinio model} \label{chap:NJL}

We apply the projection operator method to the two-flavors and three-colors 
Nambu-Jona-Lasinio (NJL) model in the chiral limit and 
derive the linear Langevin equation for the critical 
fluctuations in the chiral phase transition at finite temperature and density.
The behaviors of the critical fluctuations are different between 
above and below the critical temperature $T_c$.
In this study, we limit our discussion to the case of $T > T_c$.

The NJL Hamiltonian at finite chemical potential is
\cite{ref:HK-PR,ref:HK-PR2} 
\begin{eqnarray}
H &=& H_{0} + H_{I}, \\
H_{0} &=& \int d^3 {\bf x}
          \bar{q}({\bf x})(-i\vec{\gamma}\cdot \nabla 
          -\mu\gamma^0)q({\bf x}) , \\
H_{I} &=& -g\int d^3 {\bf x}\{
          (\bar{q}({\bf x})q({\bf x}))^2
          +(\bar{q}({\bf x})i\gamma_{5}\tau q({\bf x}))^2
          \}, \nonumber \\
\end{eqnarray}
where $H_{0}$ and $H_{I}$ are the nonperturbative Hamiltonian and 
the interaction Hamiltonian, respectively.
Here, we introduce the Pauli matrices $\tau^{i}$ ($i=1,2,3$).
We set the coupling $g = 5.01$ GeV${}^{-2}$ 
and the three dimensional cutoff $\Lambda = 650$ MeV 
so as to reproduce the pion decay constant $f_{\pi} = 93$ MeV 
and the chiral condensate $\langle \bar{q}q \rangle = (-250~{\rm MeV})^3$ 
in the chiral limit.
In particular, we are interested in the scalar channel.
Thus, we ignore the pseudoscalar part of the interaction Hamiltonian 
in the following calculation.

The quark fields $q({\bf x},t)$ and $\bar{q}({\bf x},t)$ 
are expanded as 
\begin{eqnarray}
q({\bf x},t) 
&=& \frac{1}{\sqrt{V}}
\sum_{{\bf p},s}[b_{{\bf p},s}u({\bf p},s)e^{i{\bf px}}e^{-iE_{p}t}
+d^{\dagger}_{{\bf p},s}v({\bf p},s)e^{-i{\bf px}}e^{iE_{p}t}], \nonumber \\
\\
\bar{q}({\bf x},t)
&=& \frac{1}{\sqrt{V}}
\sum_{{\bf p},s}[b^{\dagger}_{{\bf p},s}\bar{u}({\bf p},s)e^{i{\bf px}}e^{-iE_{p}t}
+d_{{\bf p},s}\bar{v}({\bf p},s)e^{-i{\bf px}}e^{iE_{p}t}], \nonumber \\
\end{eqnarray}
where $E_{p} = |{\bf p}|$.
The normalization conditions of the wave functions $u({\bf p},s)$ and 
$v({\bf p},s)$ are
\begin{eqnarray}
&& u^{\dagger}({\bf p},s)u({\bf p},s')
= v^{\dagger}({\bf p},s)v({\bf p},s') = \delta_{s,s'},\\
&& \bar{v}({\bf p},s)u({\bf p},s)
= v^{\dagger}({\bf p},s)u(-{\bf p},-s') = 0.
\end{eqnarray}
The corresponding commutation relations are given by
\begin{eqnarray}
&& [b_{{\bf p},s},b^{\dagger}_{{\bf p}',s'}]_+ 
= [d_{{\bf p},s},d^{\dagger}_{{\bf p}',s'}]_+ = \delta_{s,s'}\delta^{(3)}_{\bf p,p'}, \\
&& [q({\bf x},t),\bar{q}({\bf y},t)]_{+} = \gamma^{0}\delta^{(3)}({\bf
x}-{\bf y}), \\
&& [q({\bf x},t),q({\bf y},t)]_{+} 
= [\bar{q}({\bf x},t),\bar{q}({\bf y},t)]_{+} = 0,
\end{eqnarray}
where $[~~]_+$ means the anticommutator.

Now, we choose gross variables to define the Mori projection operator.
As is discussed in Sec. \ref{chap:Mori}, 
there are three possible candidates for gross variables:(i) order parameters 
(ii) density variables associated with conserved quantities and 
(iii) their products.
For simplicity, we ignore (iii) in this study.
Then, we have the order parameter $\bar{q}q$, 
the number density $\bar{q}\gamma^0 q$, the energy density $T^{00}$ and 
the momentum density $T^{0i}$ 
as gross variables from the conditions (i) and (ii).
However, we assume that near the critical temperature of the phase transition, 
the order parameters are much slower than 
the other density variables because of the 
critical slowing down where the relaxation time of the critical fluctuations 
indefinitely increases
(At high density, the number density can be another gross variable 
as in the glass transition. We will discuss this point in Sec. \ref{chap:Sum}.).
Thus, we exclude (ii) from our gross variables.
After all, the gross variable relevant to our calculation is 
the order parameter of the chiral phase transition.
Then, the Mori projection operator is 
\begin{eqnarray}
P O= \int d^3 {\bf x} d^3 {\bf x}' (O,\delta \sigma({\bf x}'))\cdot 
(\delta \sigma({\bf x}'),\delta \sigma({\bf x}''))^{-1}\cdot 
\delta \sigma({\bf x}''). \nonumber \\
\end{eqnarray}
Here, we have used the notation,
\begin{eqnarray}
\delta \sigma ({\bf x})
= \bar{q}({\bf x})q({\bf x}) - \langle \bar{q} ({\bf x}) q({\bf x}) \rangle_{eq}.
\end{eqnarray}
where $\langle \bar{q} ({\bf x}) q({\bf x})\rangle_{eq}$ is the expectation value 
in thermal equilibrium and vanishes above the critical temperature
\cite{ref:footnote1}.
The canonical correlation is 
\begin{eqnarray}
(X,Y) = \int^{\beta}_{0}\frac{d\lambda}{\beta}
\langle e^{\lambda H_{0}}Xe^{-\lambda H_{0}}Y \rangle_{0},
\end{eqnarray}
where
\begin{eqnarray}
\langle O \rangle_0 = \frac{1}{Z_0}{\rm Tr}[e^{-\beta H_{0}}O],
\end{eqnarray}
with $Z_0 = {\rm Tr}[e^{-\beta H_0}]$.
If we can solve the equation exactly, the system should thermalize 
with the total Hamiltonian $H$.
However, in the following calculation, 
we make approximation where the solution of the derived Langevin equation 
relaxes toward an equilibrium state 
consistent with the mean-field approximation where 
quarks behave massless free particles above the critical temperature.
Thus, we replace $H$ with $H_{0}$ in the definition of the 
canonical correlation (\ref{eqn:KCC}).
In this manner, we should choose the projection operator 
so as to abstract the process that we should describe.
In this sense, the projection operator method has more than 
the lowest order approximation of the perturbative expansion.

Substituting this Mori projection operator into the TC equation (\ref{eqn:TC-2}), 
we can derive the Langevin equation of the critical fluctuations of the 
order parameter in the chiral phase transition,
\begin{eqnarray}
\frac{d}{dt}\delta\sigma({\bf x},t)
&=& e^{iLt}PiL\delta\sigma({\bf x},0) + \int^{t}_{0}d\tau e^{iL(t-\tau)}
PiLQe^{iQL_{0}Q\tau}iL\delta\sigma({\bf x},0) \nonumber \\
 &&+\xi({\bf x},t).
\end{eqnarray}
The noise (third) term is determined so as to satisfy the 2nd F-D theorem.
And, as is shown in Appendix \ref{app:stream}, the streaming term vanishes.
Thus, in the following, we discuss the calculation 
of the memory function.

The calculation of the memory term is quite intricate because 
we must calculate the coarse-grained time evolution operator $e^{iQL_{0}Qt}$ 
instead of the normal time evolution operator $e^{iL_{0}t}$ that 
appears in normal calculations without coarse-grainings.
First, we expand the memory term with the help of the normal time evolution 
operator:
\begin{eqnarray}
\int^t_0 d\tau e^{iL(t-\tau)}PiLQ e^{iQL_0 Q\tau}  
iL \delta \sigma({\bf x})
= \int^t_0 d\tau e^{iL(t-\tau)}PiL e^{iL_0 \tau} {\mathcal B} (\tau,0) 
Q iL \delta \sigma({\bf x}), \nonumber \\
\end{eqnarray}
where
\begin{eqnarray}
{\mathcal B}(t,t_0)
&=& 1 + \sum^{\infty}_{n=1}(-i)^n \int^{t}_{t_{0}}dt_1 \cdots \int^{t_{n-1}}_{t_0}
dt_n 
\breve{L}^P_{0}(t_1-t_0)\breve{L}^P_{0}(t_{2}-t_0) \cdots 
\breve{L}^P_{0}(t_{n}-t_0), \nonumber \\
\end{eqnarray}
with
\begin{eqnarray}
\breve{L}^P_0 (t) &\equiv& e^{-iL_0 t}PL_0 e^{iL_0 t}.
\end{eqnarray}
Each term does not have the coarse-grained time evolution operator 
and is calculated exactly.
Finally, we find the n-th order term given by 
\begin{eqnarray}
\lefteqn{
\int^t_0 d\tau e^{iL(t-\tau)}PiLe^{iL_0 \tau}(-i)^n \int^{\tau}_0 dt_1 \cdots dt_n
\breve{L}^P_0 (t_1)\cdots \breve{L}^P_0 (t_n) QiL 
\delta \sigma({\bf x}) } && \nonumber \\
&=& (-1)^n \int^t_0 d\tau e^{iL(t-\tau)}PiL \int^{\tau}_0 dt_1 \cdots dt_n \nonumber \\
&&\times \frac{d^2 X_{t_n}({\bf x},{\bf x}_{n})}{dt^2_n} \cdot
\frac{d X_{t_{n-1}-t_n}({\bf x}_{n},{\bf x}_{n-1})}{dt_{n-1}}
\cdots
\frac{d X_{t_1 -t_2}({\bf x}_2,{\bf x}_1)}{dt_1}\cdot 
\delta \sigma_0 ({\bf x}_1,\tau-t_1),
\nonumber \\
\end{eqnarray}
where  $\delta \sigma_0 ({\bf x},t) = e^{iL_0 t}\delta \sigma ({\bf x},0)$ and 
\begin{eqnarray}
X_{t}({\bf x},{\bf x}'')
&=& (\delta \sigma_{0}({\bf x},t),\delta \sigma({\bf x}'))
\cdot (\delta \sigma({\bf x}'),\delta \sigma({\bf x}''))^{-1} \nonumber \\
&=& \int d^3 {\bf x}' \chi_{t}({\bf x-x'})\chi^{-1}_{0}({\bf x'-x''}).
\end{eqnarray}
In this expression, we dropped the sign of the integral for the repeated 
space variables for simple notation.
In the following, we use this notation without notice.
The concrete form of $\chi_{t}({\bf x})$ is given later.
Thus, the memory term is given by 
\begin{eqnarray}
\lefteqn{\int^t_0 d\tau e^{iL(t-\tau)}PiL e^{iL_0 \tau} {\mathcal B} (\tau,0) 
Q iL \delta \sigma({\bf x}) } && \nonumber \\
&=& \int^t_0 d\tau 
(\frac{d^2}{d\tau^2}X_{\tau}({\bf x},{\bf x}_{1})
\cdot \delta \sigma({\bf x}_{1},t-\tau)
+ e^{iL(t-\tau)}PiL_{I}\frac{d}{d\tau}\delta \sigma_0({\bf x},\tau)) 
+ \sum^{\infty}_{n=1}\int^t_0 d\tau  (-1)^n \nonumber \\
&& \times \int^{\tau}_0 d^3 {\bf x}_1 dt_1 \cdots d^3 {\bf x}_n dt_n
\frac{d^2 X_{t_n}({\bf x},{\bf x}_n)}{dt_n^2} 
\cdot
\frac{d X_{t_{n-1}-t_n}({\bf x}_n,{\bf x}_{n-1})}{dt_{n-1}} \cdots 
\frac{d X_{t_1-t_2}({\bf x}_2,{\bf x}_1)}{dt_{1}} \nonumber \\
&&\cdot (\frac{d}{d\tau}X_{\tau-t_{1}}({\bf x}_1,{\bf y})\cdot 
\delta \sigma({\bf y},t-\tau) + e^{iL(t-\tau)}PiL_{I}
\delta \sigma_0 ({\bf x}_{1},\tau-t_1)).
\label{eqn:MT-1}
\end{eqnarray}
Here, we used 
$e^{iL(t-\tau)} \delta \sigma({\bf x})= \delta \sigma({\bf x},t-\tau)$.

This expression is simplified by using the Laplace transform.
First, we ignore the terms including the interaction $iL_{I}$.
The Laplace transform of the first term is 
\begin{eqnarray}
{\mathcal L}\{ \frac{d^2}{dt^2}X_t ({\bf x},{\bf x}_1) 
\otimes \delta \sigma({\bf x}_1,t)\} = \ddot{X}^L_s({\bf x},{\bf x}_1)
\cdot \delta \sigma^L ({\bf x}_1,s), \nonumber \\
\end{eqnarray}
where ${\mathcal L}\{~~\}$ means the Laplace transform and 
\begin{eqnarray}
{\mathcal L} \{ A \otimes B \} 
&=& {\mathcal L} \{ \int^{t}_{0}d\tau A(t-\tau)B(\tau) \}.
\end{eqnarray}
Here, we introduced the following expressions,
\begin{eqnarray}
{\mathcal L}\{ \sigma({\bf x},t) \} 
&=& \int^{\infty}_{0}dt e^{-st}\sigma({\bf x},t) 
\equiv \sigma^{L}({\bf x},s), \\
{\mathcal L} \{ \frac{d}{dt} X_t ({\bf x},{\bf x}_1) \}
&=& \int^{\infty}_{0}dt e^{-st}\frac{d}{dt} X_t ({\bf x},{\bf x}_1) 
\equiv \dot{X}^L_s({\bf x},{\bf x}_1), \\
{\mathcal L} \{ \frac{d^2}{dt^2} X_t ({\bf x},{\bf x}_1) \}
&=& \int^{\infty}_{0}dt e^{-st}\frac{d^2}{dt^2}X_t ({\bf x},{\bf x}_1) 
\equiv \ddot{X}^L_s({\bf x},{\bf x}_1). 
\end{eqnarray}
The Laplace transform of the second term is 
\begin{eqnarray}
&& {\mathcal L} 
\{ -\int^t_0 dt_1 \frac{d^2 X_{t_1}({\bf x},{\bf x}_1)}{dt_1^2}\cdot 
\frac{dX_{t-t_1}({\bf x}_1,{\bf y})}{dt} \otimes \delta \sigma({\bf y},t) \}   
\nonumber \\
&& = -\ddot{X}^L_{s}({\bf x},{\bf x}_1) \cdot 
\dot{X}^L_{s}({\bf x}_1,{\bf y}) \cdot \delta \sigma^L ({\bf y},s).
\end{eqnarray}
All other terms can be calculated in the same way.
Thus, the sum of all the Laplace transform of the memory term 
is given by 
\begin{eqnarray}
&& {\mathcal L} \{ \int^t_0 ds 
\frac{d^2}{ds^2}X_{s}({\bf x},{\bf x}_{1})\cdot \delta \sigma({\bf x}_{1},t-s)
+ \sum^{\infty}_{n=1}  (-1)^n 
\int^{t}_{0}d\tau \int^{\tau}_0 dt_1 \cdots dt_n \nonumber \\
&&\times \frac{d^2 X_{t_n}({\bf x},{\bf x}_n)}{dt_n^2} 
\cdots 
\frac{d X_{t_1-t_2}({\bf x}_2,{\bf x}_1)}{dt_{1}} 
\cdot \frac{d}{d\tau}X_{\tau-t_{1}}({\bf x}_1,{\bf y})\cdot 
\delta \sigma({\bf y},t-\tau) \} \nonumber \\
&&= \int \frac{d^3 {\bf k}}{(2\pi)^3} e^{i{\bf kx}} 
\ddot{\chi}^L_s({\bf k})/(\chi_{0}({\bf k}) + \dot{\chi}^L_s ({\bf k})).
\end{eqnarray}
Here, we used that  
\begin{eqnarray}
\lefteqn{\chi_t ({\bf k})} && \nonumber \\
&=& 
\int d^3 ({\bf x-x'}) e^{-i{\bf k(x-x')}}
(\delta \sigma_0 ({\bf x},t),\delta \sigma({\bf x}',0))
\nonumber \\
&=& \frac{N_{c}N_{f}}{\beta V}
\sum_{\bf p}
\{
(
n^+(E_{\bf k+p}) - n^+(E_{\bf p}) + n^-(E_{\bf k+p}) - n^-(E_{\bf p}) 
) \nonumber \\
&& \times \frac{E_{\bf p}E_{\bf p+k} - {\bf p(p+k)} }
{E_{\bf p}E_{\bf p+k}(E_{\bf p} - E_{\bf p+k})} e^{i(E_{\bf p}-E_{\bf p+k})t}  \nonumber \\
&& + (
1- n^+(E_{\bf p}) - n^-(E_{\bf p+k})
) 
\frac{E_{\bf p}E_{\bf p+k} + {\bf p(p+k)} }
{E_{\bf p}E_{\bf p+k}(E_{\bf p} + E_{\bf p+k})} \nonumber \\
&& \times (e^{i(E_{\bf p} + E_{\bf p+k})t} + e^{-i(E_{\bf p} + E_{\bf p+k})t})
\}, 
\end{eqnarray}
where $N_{c} = 3$, $N_f = 2$, $E_{\bf k} = |{\bf k}|$ and 
\begin{eqnarray}
n^{\pm}(E) = \frac{1}{e^{\beta(E\mp \mu)}+1}.
\end{eqnarray}
The Laplace transforms are defined by
\begin{eqnarray}
\dot{\chi}^{L}_{s}({\bf k}) = \int^{\infty}_{0} dt e^{-st} 
\frac{d}{dt}\chi_t({\bf k}), \\
\ddot{\chi}^{L}_{s}({\bf k}) = \int^{\infty}_{0} dt e^{-st} 
\frac{d^2}{dt^2}\chi_t({\bf k}).
\end{eqnarray}
This term has the well-known structure and can be interpreted in 
the kinematical way.
The first term is the scattering process and the second term 
is pair creation and annihilation processes.

\begin{figure}\leavevmode
\begin{center}
\epsfxsize=6cm
\epsfbox{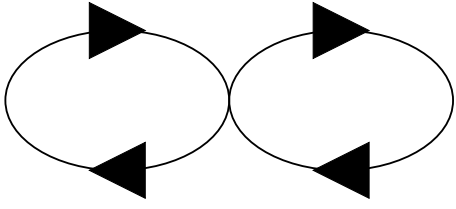}
\caption{The ring diagram that contributes the calculation of the memory term.}
\label{fig:RPA}
\end{center}
\end{figure}

Next, we calculate the terms including $iL_{I}$.
The uncalculating factor $PiL_{I}\delta \sigma_{0}({\bf x},t)$ in Eq. (\ref{eqn:MT-1})
contains various contributions.
Here, we are concerned with 
the contribution of 
the particle-antiparticle loop diagram shown in Fig. \ref{fig:RPA}.
(Then, our Langevin equation shows consistent behaviors 
with the equilibrium results in the mean-field approximation, 
as we will see later.)
In this random phase approximation, we can calculate the interaction part 
by using Wick's theorem,
\begin{eqnarray}
\lefteqn{
( iL_{I}\delta \sigma_{0}({\bf x},\tau), 
\delta \sigma({\bf z}) ) }&& \nonumber \\
&\sim& -ig\int d^3 {\bf y}
2[
\contradd{q_{0}^{\alpha}({\bf y},-i\lambda)}
{\bar{q}^{\gamma}_{0}({\bf x},\tau-i\lambda)}
\hspace{2.2cm}
\contradd{\bar{q}^{\alpha}_{0}({\bf y},-i\lambda)}
{q^{\gamma}_{0}({\bf x},\tau-i\lambda)} \hspace{2.2cm}
\contradd{q_{0}^{\beta}({\bf y},-i\lambda)}
{\bar{q}^{\delta}({\bf z})}
\hspace{0.8cm}
\contradd{\bar{q}^{\beta}_{0}({\bf y},-i\lambda)}
{q^{\delta}({\bf z})}
\hspace{0.8cm} \nonumber \\
\nonumber \\
&& - 
\contraddd{\bar{q}^{\gamma}_{0}({\bf x},\tau-i\lambda)}
{q^{\beta}_{0}({\bf y},-i\lambda)}
\hspace{1.8cm}
\contraddd{q_{0}^{\gamma}({\bf x},\tau-i\lambda)}
{\bar{q}^{\beta}_{0}({\bf y},-i\lambda)} \hspace{1.8cm}
\contradd{q_{0}^{\alpha}({\bf y},-i\lambda)}
{\bar{q}^{\delta}({\bf z})}
\hspace{0.9cm}
\contradd{\bar{q}^{\alpha}_{0}({\bf y},-i\lambda)}
{q^{\delta}({\bf z})}
\hspace{0.9cm}
]
\nonumber \\
&=& 
-2g\beta \int d^3 {\bf y} 
\frac{d}{d\tau} \chi_{\tau}({\bf x},{\bf y}) \chi_{0}({\bf y},{\bf z}).
\end{eqnarray}
In the second line of the r. h. s. of the equation, 
we approximate the function by calculating the contribution of the 
ring diagram shown in Fig.~\ref{fig:RPA}.
Here, we introduced the contraction defined in Appendix \ref{app:Propa}.
Finally, the Laplace transform of the remaining part of the memory term is given by,
\begin{eqnarray}
&&{\mathcal L}\{\int^t_0 d\tau 
e^{iL(t-\tau)}PiL_{I}\frac{d}{d\tau}\delta \sigma_0({\bf x},\tau)
+ \sum^{\infty}_{n=1}\int^t_0 d\tau  (-1)^n \nonumber \\
&& \times \int^{\tau}_0 dt_1 
\cdots dt_n \frac{d^2 X_{t_n}({\bf x},{\bf x}_n)}{dt_n^2} 
\cdots 
\frac{d X_{t_1-t_2}({\bf x}_2,{\bf x}_1)}{dt_{1}} 
\cdot e^{iL(t-\tau)}PiL_{I}\delta \sigma_0 ({\bf x}_{1},\tau-t_1))\} \nonumber \\
&=& \int \frac{d^3 {\bf k}}{(2\pi)^3}e^{i{\bf kx}}
\frac{ \ddot{\chi}^L_s ({\bf k}) }{(\chi_{0}({\bf k}) + \dot{\chi}^L_s({\bf k}))} 
(-2g\beta\chi_{0}({\bf k}))\delta \sigma^L({\bf k},s). \nonumber \\
\end{eqnarray}
Summarizing the above results, the memory term is given by 
\begin{eqnarray}
\int^t_0 d\tau e^{iL(t-\tau)}PiLQ e^{iQL_0 Q\tau}  
iL \delta \sigma({\bf x})
= -\int^t_0 d\tau \int d^3 {\bf x}' 
\Gamma({\bf x}-{\bf x}',\tau)\delta \sigma ({\bf x}',t-\tau). \nonumber \\
\end{eqnarray}
Here, the memory function $\Gamma({\bf x},t)$ is given by 
the inverse-Laplace transform of $\Gamma^L ({\bf x},s)$,
\begin{eqnarray}
\Gamma^L ({\bf x},s) =
\int \frac{d^3 {\bf k}}{(2\pi)^3}e^{i{\bf kx}}
\frac{ -\ddot{\chi}^L_s ({\bf k}) }{(\chi_{0}({\bf k}) + \dot{\chi}^L_s({\bf k}))}
(1-2g\beta\chi_{0}({\bf k})). \nonumber \\
\label{eqn:MF}
\end{eqnarray}
In short, the memory function that includes the coarse-grained time evolution 
operator $e^{iQL_0 Qt}$ is expressed by the combination of the 
normal correlation function $\chi_{t}({\bf k})$.
This fact was pointed out in 
\cite{ref:Sawada} for the first time.

Finally, we discuss the noise term.
In principle, the noise term also can be calculated by using 
the perturbative expansion.
However, we do not need the concrete form of the noise 
in the following calculations.
Thus, we determine the noise so as to reproduce the 2nd F-D theorem.
Then, the equation is given by
\begin{eqnarray}
\frac{d}{dt} \delta \sigma({\bf k},t) 
=
-\int^{t}_{0}d\tau \Gamma ({\bf k},\tau)\delta \sigma({\bf k},t-\tau) 
+ \xi ({\bf k},t) \label{eqn:LLE},
\end{eqnarray}
where the introduced noise $\xi({\bf k},t)$ has 
the first and second order correlations given by 
\begin{eqnarray}
(\xi({\bf k},t),\delta \sigma({\bf k}',0))
&=& 
\langle \xi({\bf k},t) \rangle_{0} = 0,\\
(\xi({\bf k},t),\xi({\bf k}',t'))
&=& V \delta^{(3)}_{\bf k,k'}
\Gamma({\bf k},t-t')\chi_{0}({\bf k}).
\end{eqnarray}
The above equation results from the naive application of the 
projection operator method.
However, we cannot interpret this equation as a Langevin equation.
To derive the Langevin equation, we must employ the renormalization of the 
memory function, as we will discuss in Sec. \ref{chap:renor}.

\section{Renormalized Langevin equation} \label{chap:renor}

\begin{figure}\leavevmode
\begin{center}
\epsfxsize=8cm
\epsfbox{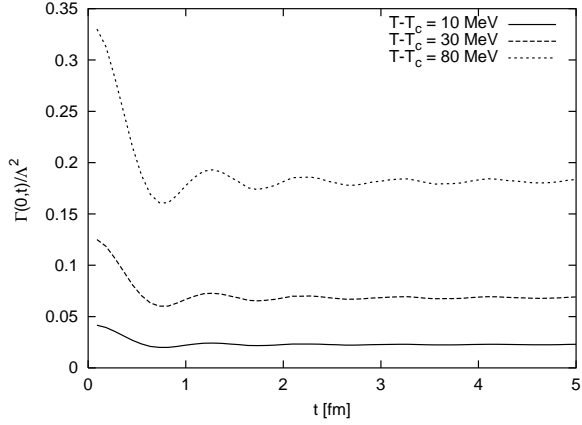}
\caption{The time dependence of the memory function at $\mu = 200$ MeV.
The solid, dashed and dotted lines are the memory function for 
temperatures $T-T_c = 10$ MeV, $30$ MeV and $80$ MeV.}
\label{fig:TD_Total}
\end{center}
\end{figure}

We have applied the projection operator method to the NJL model 
and derived the equation (\ref{eqn:LLE}).
Unfortunately, we cannot regard this equation as a Langevin equation, 
because of the problem of the long time correlation of the 
memory function.
In Fig. \ref{fig:TD_Total}, the time dependence of the memory function 
$\Gamma({\bf 0},t)$ is shown at $\mu = 200$ MeV.
The solid, dashed and dotted lines are the memory function for 
temperatures $T-T_c = 10$ MeV, $30$ MeV and $80$ MeV.
The memory functions converge to finite values at late time.
This behavior is unphysical because the memory function is given by the 
time correlation function of the noise from the 2nd F-D theorem.
It is clear from the definition of the noise (\ref{eqn:TC-1}) that 
because of the projection operator $Q$, all gross variables are 
excluded from the time evolution of the noise and hence 
the time correlation must converge to zero rapidly.

The long time correlation means that 
a variable associated with a long time scale is still 
included in the noise because of the incompleteness of the definition 
of the Mori projection operator
\cite{ref:Fick}.
As an example, we consider an exactly solvable model
(See Appendix \ref{app:Brown} for details.).
The Hamiltonian is 
\begin{eqnarray}
H = \frac{p^2}{2M} + \frac{M\omega_{0}^2}{2}x^2 
+ \sum_{i}[
\frac{p_{i}^2}{2} + \frac{\omega_{i}^2}{2}
(x_{i}-\frac{\gamma_{i}}{\omega_{i}^2}x)^2
].
\end{eqnarray}
It is known that we must choose two gross variables $p$ and $x$ 
to derive the correct Langevin equation.
Then, we have 
\begin{eqnarray}
\frac{d}{dt} x(t) &=& \frac{p(t)}{M}, \\
\frac{d}{dt} p(t) &=& -M\omega^2_0 x(t) - \int^{t}_0 ds \Xi(t-s) p(s) + f(t), \nonumber \\
\label{eqn:Ex-1}
\end{eqnarray}
where the memory function $\Xi(t)$ and the noise $f(t)$ are given by
\begin{eqnarray}
\Xi(t) 
&=& (f(t),f(0))\cdot (p,p)^{-1} 
= \frac{1}{M}\sum^{N}_{j=1}\frac{\gamma_i}{\omega^2_j}\cos \omega_j (t-t'), 
\nonumber \\
\\
f(t)
&=& \sum^{N}_{j=1}\gamma_i[ \cos \omega_j t 
(x_i(0) - \frac{\gamma_j}{\omega^2_j}x(0)) + \frac{\sin \omega_j t}{\omega_j}p_j (0) ], \nonumber \\
\end{eqnarray}
respectively.
On the other hand, when we choose only one gross variable $p$, 
the resulting equation has the long time correlation,
\begin{eqnarray}
\frac{d}{dt}p(t) = - \int^{t}_{0}ds \Xi'(t-s) p(s) + f'(t), \label{eqn:Ex-2}
\end{eqnarray}
where
\begin{eqnarray}
\Xi'(t) &=& (f'(t),f'(0))\cdot (p,p)^{-1}
= \omega^2_0 + \Xi(t), \\
f'(t) &=& -M \omega^2_0 x(0) + f(t).
\end{eqnarray}
As is discussed in Appendix \ref{app:Brown}, if we observe the system with 
long time scale, $\Xi (t)$ converges to zero quickly and does not have 
long time correlation.
However, $\Xi'(t)$ does not converge to zero because of the frequency $\omega^2_0$ 
and has long time correlation.
As just described, the behavior of the memory function strongly depends on 
the definition of the Mori projection operator.
In order to obtain the memory function without the long time correlation, 
we must prepare the complete set of gross variables.

\begin{table}\leavevmode
\begin{center}
\begin{tabular}{c|ccccc}
Model & GV1 & GV2 & MF1 & MF2 & F \\ \hline
\hline
Exact  & $p$ & $x$ & $\Xi'(t)$ & $\Xi(t)$ & $\omega^2_0$ \\
NJL & $\delta \sigma$ & ?  &  $\Gamma({\bf k},t)$ & $\Phi({\bf k},t)$ 
& $\Omega^2({\bf k},t)$ \\ 
\end{tabular}
\caption{The correspondence between the exact model and the Langevin equation 
in the NJL model.
The first two columns indicate the gross variables (GV) in each model.
The next two columns denote the memory function with long time correlation (MF1) 
and that without long time correlation (MF2).
The last column shows the frequency function (F).
}
\label{tab:correspondence}
\end{center}
\end{table}

The correspondence of the exact model and our equation is given 
in TABLE \ref{tab:correspondence}.
The gross variable $p$ in the exact model corresponds to 
$\delta \sigma({\bf k},t)$ in our Langevin equation.
Therefore, in order to derive the Langevin equation without long time correlation, 
we must find out another variable corresponding to $x$ in our model and 
redefine the Mori projection operator with the two gross variables.
However, another gross variable in the NJL model is not obvious 
because the NJL Hamiltonian is composed of $q$ and $\bar{q}$, and 
does not include the composite operator $\sigma = \bar{q}q$ explicitly.

\begin{figure}\leavevmode
\begin{center}
\epsfxsize=8cm
\epsfbox{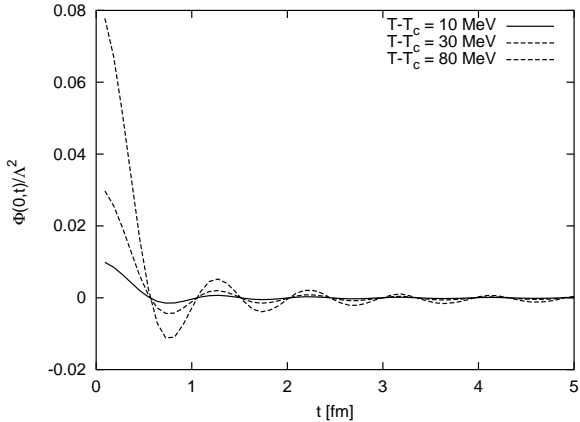}
\caption{The time dependence of the renormalized memory function at $\mu =200$ MeV.
The solid, dashed and dotted lines are the memory function for 
temperatures $T-T_c = 10$ MeV, $30$ MeV and $80$ MeV.}
\label{fig:Renor}
\end{center}
\end{figure}

\begin{figure}\leavevmode
\begin{center}
\epsfxsize=8cm
\epsfbox{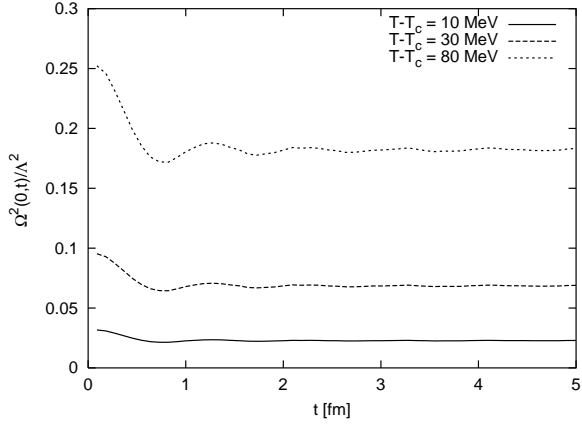}
\caption{The time dependence of the frequency function at $\mu =200$ MeV.
The solid, dashed and dotted lines are the memory function for 
temperatures $T-T_c = 10$ MeV, $30$ MeV and $80$ MeV.}
\label{fig:Fre1}
\end{center}
\end{figure}

Therefore, in this paper, we derive the correct memory function in another way.
First, we assume that there exists the exact correspondence of 
the exact model with our equation, and speculate the correct form of the 
memory function.
By comparison Eq. (\ref{eqn:Ex-2}) with Eq. (\ref{eqn:LLE}), 
our memory function corresponds to $\Xi'(t)$, while 
the correct memory function should be given by the counterpart of $\Xi (t)$.
In order to obtain the correct memory function in our model, we should 
separate the $\Gamma({\bf k},t)$ into two parts: 
the frequency part and the dissipation part.
In quantum field theory, the frequency is given by the real part of the 
self energy in energy-momentum space, 
that is, the imaginary part of the memory function.
Thus, we separate the imaginary part from the memory function to 
redefine the memory function without the long time tail.
As a result, our memory function is separated into the following two parts,
\begin{eqnarray}
\Gamma({\bf k},t) &=& \Omega^{2}_{\bf k}(t) + \Phi({\bf k},t), \label{eqn:reno-con1}\\
\Omega^{2}_{\bf k}(t) 
&=& \int \frac{d\omega}{2\pi} 
i{\rm Im}[\Gamma^{L} ({\bf k},-i\omega+\epsilon)]
e^{-i\omega t}, \label{eqn:reno-con2}\\
\Phi({\bf k},t) 
&=& \int \frac{d\omega}{2\pi} 
{\rm Re}[\Gamma^{L}({\bf k},-i\omega+\epsilon)] 
e^{-i\omega t}.\label{eqn:reno-con3}
\end{eqnarray}
The memory function without long time correlation is given by 
the "renormalized" memory function $\Phi({\bf k},t)$, 
while $\Omega^{2}_{\bf k}(t)$ gives the frequency function.

The time dependence of the renormalized memory function and the frequency function 
at $\mu = 200$ MeV 
are given in Figs. \ref{fig:Renor} and \ref{fig:Fre1}, respectively.
The solid, dashed and dotted lines are the memory function for 
temperatures $T-T_c = 10$ MeV, $30$ MeV and $80$ MeV.
The renormalized memory function relaxes rapidly and vanishes at late time, 
while the frequency function converges to a finite value depending on the 
temperature.
In this manner, we can get rid of the long time correlation of the 
memory function by renormalization.

Consequently, the correlation properties of the noise are also changed.
The correct noise is related not to the memory function 
but to the renormalized memory function 
by the 2nd F-D theorem.
At last, we have obtained the renormalized linear Langevin equation, 
\begin{eqnarray}
\frac{d}{dt} \delta \sigma({\bf k},t) 
= -\int^{t}_{0}d\tau \Omega^2_{\bf k} (\tau) \delta \sigma({\bf k},t-\tau) 
-\int^{t}_{0}d\tau \Phi 
({\bf k},\tau)\delta \sigma({\bf k},t-\tau) 
+ \xi ({\bf k},t), \nonumber \\
\label{eqn:RLLE}
\end{eqnarray}
where
\begin{eqnarray}
( \xi({\bf k},t), \delta \sigma({\bf k}',0) )
&=& \langle \xi({\bf k},t) \rangle = 0, \\
(\xi({\bf k},t),
\xi^{\dagger}({\bf k'},t'))
&=& V\delta^{(3)}_{\bf k,k'} \Phi({\bf k},t-t')\chi_{0}({\bf k}).
\end{eqnarray}

From the behavior of the renormalized memory function, we 
can see the typical time scale of the noise is given by about $1$ fm, 
as is shown in Fig. \ref{fig:Renor}.
As is discussed above, the time scale of the renormalized memory function 
is nothing but that of the noise.
Thus, the typical time scale of 
the microscopic variables we have coarse-grained is about $1$ fm.
In our Langevin dynamics approach, the time scale of gross variables 
must be larger than that of the coarse-grained variables.
Thus, our Langevin equation is reliable when 
the relaxation time of $\delta \sigma({\bf k},t)$ is larger than $1$ fm.
This condition is always satisfied near the critical points 
because of the critical slowing down.

This Langevin equation is the equation of operators, that is, 
the quantum Langevin equation.
However, in the following discussions, we solve Eq. (\ref{eqn:RLLE}) 
as a semiclassical equation with a noise as a classical random field, 
because the concept of the noise as an operator is not clear.
For this purpose, we introduce a classical noise that reproduces 
the correlation properties defined above, 
\begin{eqnarray}
\ll \xi({\bf k},t)\delta \sigma({\bf k}',0) \gg &=& \ll \xi({\bf k},t) \gg
= 0, \\
\ll \xi({\bf k},t)\xi^{*}({\bf k'},t') \gg 
&=& (\xi({\bf k},t),
\xi^{\dagger}({\bf k'},t')), \label{eqn:flu-corre2}
\end{eqnarray}
where $\ll~~\gg$ means the average for the noise with a suitable 
stochastic weight.
However, even after this replacement, 
the quantum effect is still included in $\Omega^2({\bf k},t)$, 
$\Phi({\bf k},t)$ and $\chi_0({\bf k})$.
In this sense, this is a semiclassical Langevin equation.

The correlations of the noise obtained by applying renormalization 
is consistent with those in the generalized Langevin equation for 
the Brownian motion discussed in 
\cite{ref:Kubo-book}, where 
the correlation of the noise is given by the real part of the memory function 
rather than the memory function itself as is discussed above.
Otherwise, the Langevin equation does not show thermalization at late time.

\begin{figure}\leavevmode
\begin{center}
\epsfxsize=8cm
\epsfbox{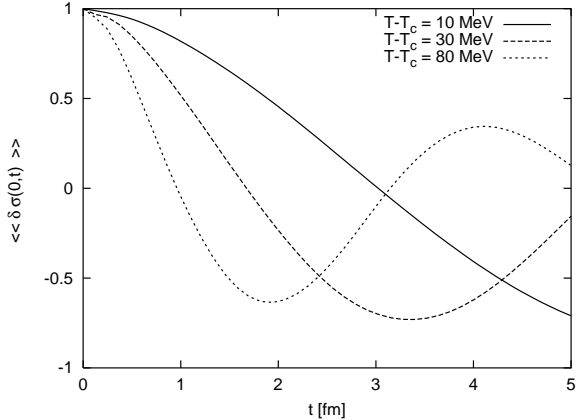}
\caption{The time evolution of $\ll \delta \sigma({\bf 0},t) \gg$ 
at $\mu = 200$ MeV. 
The solid, dashed and dotted lines are the $\ll \delta \sigma({\bf 0},t) \gg$
for temperatures $T-T_c = 10$ MeV, $30$ MeV, $80$ MeV.}
\label{fig:OP}
\end{center}
\end{figure}

The averaged time evolution of the critical fluctuations of the order parameter 
at vanishing momentum is shown in Fig.~\ref{fig:OP}.
One can see that the nonequilibrium fluctuations relax exhibiting oscillation 
and finally converge to zero.
This indicates that the critical dynamics of the chiral phase transition 
may not be described by a simple diffusion-type equation like the time-dependent 
Ginzburg-Landau (TDGL) equation.
The relaxation time of the critical fluctuations becomes larger as 
the temperature approaches toward the critical temperature because of the 
critical slowing down that is investigated carefully in Sec. \ref{chap:CSD}.

\section{Thermalization and power spectrum} \label{chap:PS}

We have derived the linear Langevin equation in the projection operator method, 
employing the lowest order approximation of the perturbative expansion 
and the renormalization of the memory function.
In this section, we show that the Langevin equation converges to 
an equilibrium state consistent with the mean-field approximation 
as time goes on.

The relaxation time of the critical fluctuations diverges at the 
critical point because of the critical slowing down.
Thus, the critical temperature of the second order phase transition 
can be determined by the temperature where $\Phi({\bf 0},t)$, 
and hence $\Gamma({\bf 0},t)$, vanishes.
From Eq. (\ref{eqn:MF}), the vanishing point is given by 
the following condition,
\begin{eqnarray}
1-2g\beta \chi_0 ({\bf 0})|_{T=T_c} = 0.
\end{eqnarray}
This condition is nothing but the self-consistency condition of the 
chiral condensate calculated in the mean-field approximation
\cite{ref:KKKN3}.
Therefore, the critical temperature of second order phase transition 
in the linear Langevin equation is completely same as 
the conventional mean-field result.
Besides, 
this condition can be regarded as a kind of the Thouless criterion
\cite{ref:Thou}.
On the other hand, it is not clear that the critical temperature of the 
first order phase transition can be determined by the same condition.
However, from the numerical result discussed below, 
one can see that 
the critical slowing down occurs even for the first order phase transition.

Next, we show that the linear Langevin equation reveals thermalization.
From Fig. \ref{fig:OP}, 
$\delta \sigma ({\bf k},t)$ converges to zero at late time for arbitrary 
initial states.
This is one evidence of thermalization.
The other evidence is the correlation function of $\delta \sigma({\bf k},t)$.
From the renormalized Langevin equation, 
we can derive the equation of the correlation function,
\begin{eqnarray}
\frac{d}{dt}\ll \delta \sigma({\bf k},t) \delta \sigma(-{\bf k},0) \gg
&=& -\int^{t}_{0}d\tau \Omega^2_{\bf k} (\tau) 
\ll \delta \sigma({\bf k},t-\tau) \delta \sigma(-{\bf k},0) \gg \nonumber \\
&& -\int^{t}_{0}d\tau \Phi 
({\bf k},\tau) \ll \delta \sigma({\bf k},t-\tau) \delta \sigma(-{\bf k},0) \gg .
\nonumber \\
\end{eqnarray}
When, we prepare the thermal equilibrium state as an initial state, 
the equation represents the correlation function in equilibrium,
\begin{eqnarray}
\ll \delta \sigma({\bf k},0) \delta \sigma(-{\bf k},0) \gg 
= ( \delta \hat{\sigma}({\bf k},0), \delta \hat{\sigma}(-{\bf k},0) ) 
= \chi_0({\bf k}).
\end{eqnarray}
The solution of the differential equation is defined in $t \geq 0$.
Thus, we assume that the solution is symmetric at $t=0$.
Then, the correlation function is given by
\begin{eqnarray}
\lefteqn{\ll \delta \sigma({\bf k},t) \delta \sigma(-{\bf k},0) \gg}
&& \nonumber \\
&=&
\int^{\infty}_{-\infty}\frac{d\omega}{2\pi} 
{\rm Re} \left[\frac{2\chi_{0}({\bf k})}
{-i\omega + 
\Omega_{\bf k}^2 (\omega) + \Phi ({\bf k},\omega)} \right] 
e^{-i\omega t}\nonumber \\
&=&
\int^{\infty}_{-\infty}\frac{d\omega}{2\pi} 
\frac{2\Phi({\bf k},\omega)\chi_{0}({\bf k})}
{|-i\omega + 
\Omega_{\bf k}^2 (\omega) + \Phi ({\bf k},\omega)|^2} e^{-i\omega t},
\label{eqn:PowerS}
\end{eqnarray}
where
\begin{eqnarray}
\Omega^2_{\bf k}(\omega) 
&=& \int^{\infty}_{0}dt \Omega^2_{\bf k}(t) e^{i\omega t} 
= i{\rm Im}\Gamma^{L}({\bf k},-i\omega + \epsilon), \\
\Phi({\bf k},\omega)
&=& \int^{\infty}_{0}dt \Phi({\bf k},t) e^{i\omega t} 
= {\rm Re}\Gamma^{L}({\bf k},-i\omega + \epsilon).
\end{eqnarray}
It should be noted that $\Omega^2_{\bf k}(\omega)$ and 
$\Phi({\bf k},\omega)$ are the odd and even functions of 
$\omega$, respectively.

\begin{figure}\leavevmode
\begin{center}
\epsfxsize=8cm
\epsfbox{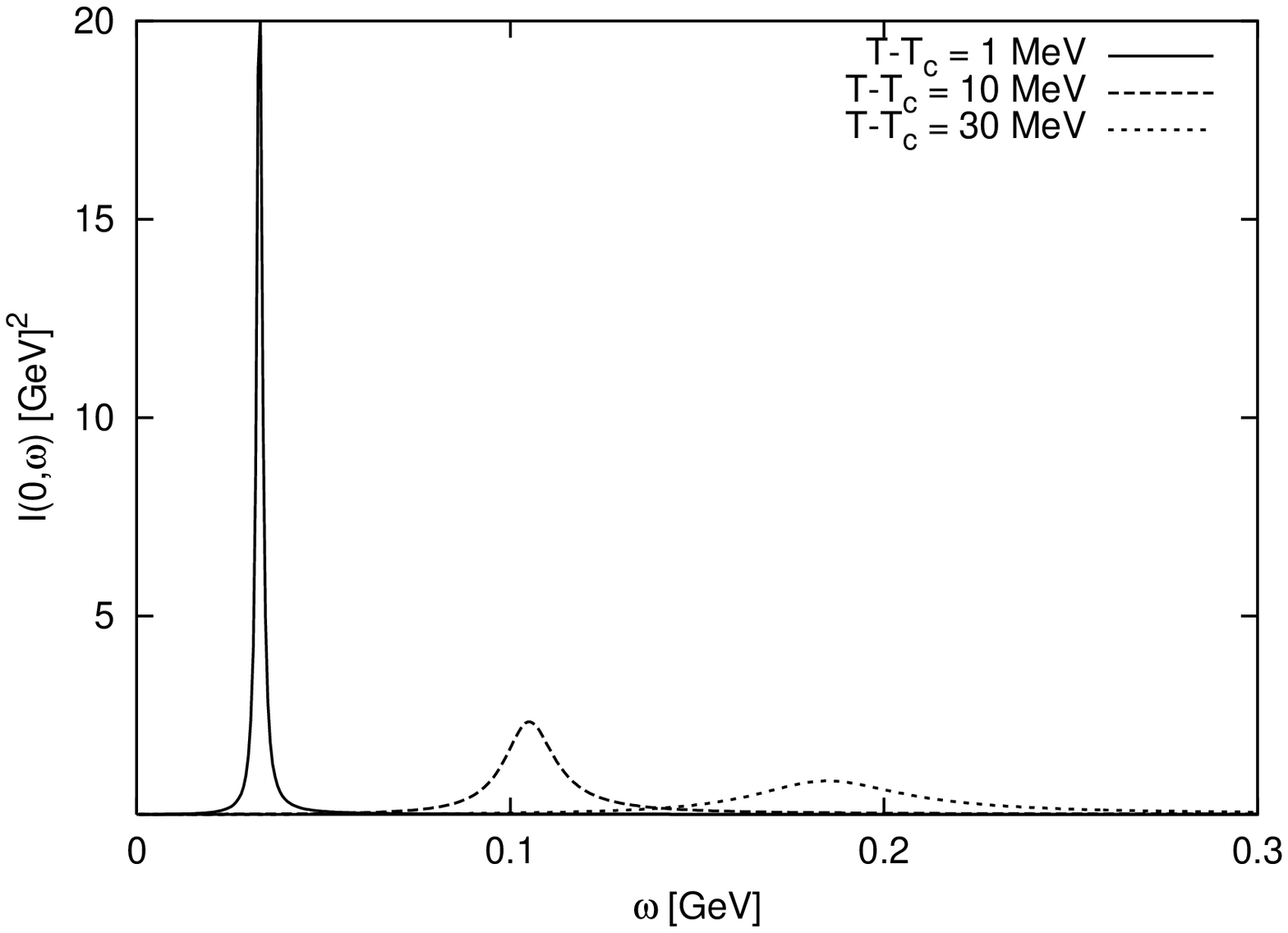}
\caption{The temperature dependence of the power spectrum at ${\bf k}={\bf 0}$ 
for a fixed chemical potentials $\mu = 0$.
The solid, dashed and dotted lines are for $T-T_c = 1$ MeV, $10$ MeV and $30$ MeV, 
respectively.}
\label{fig:PS_0}
\end{center}
\end{figure}

The same result can be obtained from the Wiener-Khinchin theorem, 
where the correlation function in equilibrium is given by 
\begin{eqnarray}
\lefteqn{ \lim_{t'\rightarrow \infty}
\ll \delta \sigma({\bf x+x'},t+t') \delta \sigma({\bf x'},t') \gg } &&
\nonumber \\
&&= \int^{\infty}_{-\infty}\frac{d\omega d^3 {\bf k}}{(2\pi)^4}
I({\bf k},\omega) e^{-i\omega t}e^{i{\bf k,x}}.
\end{eqnarray}
Here, the power spectrum $I({\bf k},\omega)$ is defined by 
\begin{eqnarray}
I({\bf k},\omega) = 
\lim_{T,V \rightarrow \infty}
\frac{1}{TV}
\ll |\delta \sigma ({\bf k},\omega)|^2 \gg, 
\end{eqnarray}
where
\begin{eqnarray}
\delta \sigma({\bf k},\omega) 
= \lim_{T \rightarrow \infty} 
\int^{T}_{0}dt \delta \sigma({\bf k},t) e^{i\omega t}.
\end{eqnarray}
Here, $T$ is the period of the time evolution of $\delta \sigma({\bf k},t)$.
In this derivation, we do not use the information of the initial state.
However, the result is completely same as Eq. (\ref{eqn:PowerS}).
By this means, we obtain the same expression of the correlation function 
in different two ways under the assumption of thermalization.
Furthermore, if we ignore the oscillation term, 
the above result reproduces the same result as the power spectrum in 
the Brownian motion
\cite{ref:Kubo-book}.
In this sense, our Langevin equation reveals thermalization.

And this is the reason that we adapted the renormalization procedure 
discussed in the previous section.
To get rid of the long time correlation of the memory function, 
it is possible to apply another condition of renormalization; 
\begin{eqnarray}
\Omega^2_{\bf k}(t) &\longrightarrow& \Gamma({\bf k},\infty), \\
\Phi({\bf k},t) &\longrightarrow& \Gamma({\bf k},t) - \Gamma({\bf k},\infty).
\end{eqnarray}
However, it is clear that we cannot show thermalization 
in this renormalization condition.
Therefore, in order to assure thermalization, we should employ 
the renormalization condition discussed in the previous section.

The temperature dependence of the power spectrum is shown in Fig.~\ref{fig:PS_0}.
We can see that the peak moving toward origin becomes prominent 
as the temperature is lowered toward $T_c$.
The power spectrum characterizes the space-time correlation in the 
energy-momentum space and hence can be interpreted as the spectral 
function in thermal field theory.
Then the peak with narrow width reveals the existence of a collective mode 
whose energy tends to vanish as the temperature approaches toward $T_c$.
Such a mode is called a soft mode.
The soft mode appears when the system becomes unstable for external perturbations
\cite{ref:HK,ref:KKKN}.
The temperature dependence of the power spectrum is consistent with 
the the previous result, where the spectral function is calculated 
in the linear response theory
\cite{ref:HK}.

\section{Enhancement of critical slowing down} \label{chap:CSD}

As is well-known, the long wave length component of the order parameter 
shows extraordinary large fluctuations near the critical point 
because of the increase of the relaxation time.
This slowing down of the fluctuations is called 
the critical slowing down
\cite{ref:vanHove}.
In this section, we discuss the behavior of the critical slowing down 
in $T$-$\mu$ plane.

\begin{figure}
\leavevmode
\begin{center}
\epsfxsize=8cm
\epsfbox{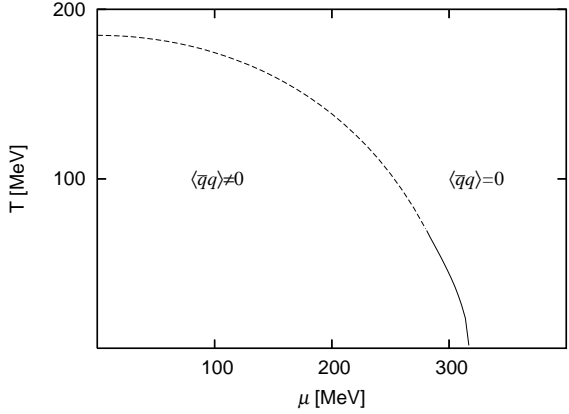}
\caption{The phase diagram in $T-\mu$ plane in the mean-field approximation.
The dashed and solid lines denote the critical line 
of the second and the first order phase transition, respectively.
The tricritical point is located at $(T,\mu) = (70,280)$ MeV.}
\label{fig:PD}
\end{center}
\end{figure}

\begin{figure}
\leavevmode
\begin{center}
\epsfxsize=8cm
\epsfbox{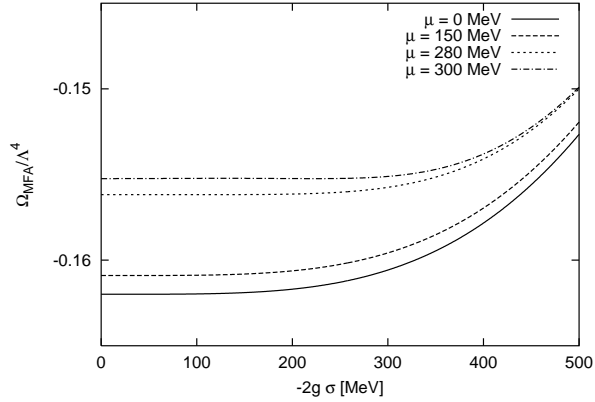}
\caption{The thermodynamic potential per unit volume on the critical line of the chiral phase transition.
The solid, dashed, dotted and dotted dashed lines represent 
the thermodynamic potential at $\mu = 0$, $150$, $280$, $300$ MeV, respectively.}
\label{fig:EP}
\end{center}
\end{figure}

For this purpose, first, we calculate the phase structure of the NJL model.
As is discussed, the renormalized Langevin equation (\ref{eqn:RLLE}) 
has the same critical temperature as that in the mean-field approximation.
Thus, we should calculate the thermodynamic potential 
in the mean-field approximation in order to fix the phase diagram, 
that is shown in Fig.~\ref{fig:PD}.
The dashed line represents the critical line of the second order phase 
transition and starts from $(T,\mu) = (185,0)$ MeV.
The order of the phase transition changes from second to first at the 
tricritical point located at $(T,\mu) = (70,280)$ MeV.
The solid line expresses the critical line 
of the first order phase transition that ends at $(T,\mu) = (0,317)$ MeV.

The corresponding thermodynamic potential per unit volume 
on the critical line is shown in Fig.~\ref{fig:EP}.
The solid, dashed, dotted and dotted dashed lines represent 
the thermodynamic potential at $\mu = 0$, $150$, $280$, $300$ MeV, respectively.
The bottom of the thermodynamic potential becomes broader as the 
chemical potential approaches toward the tricritical point.
This reflects the existence of the large fluctuations around the tricritical point.
At $\mu=300$ MeV, the order of the phase transition is first and hence 
the thermodynamic potential has two local minima, although it is invisible.

\begin{figure}\leavevmode
\begin{center}
\epsfxsize=8cm
\epsfbox{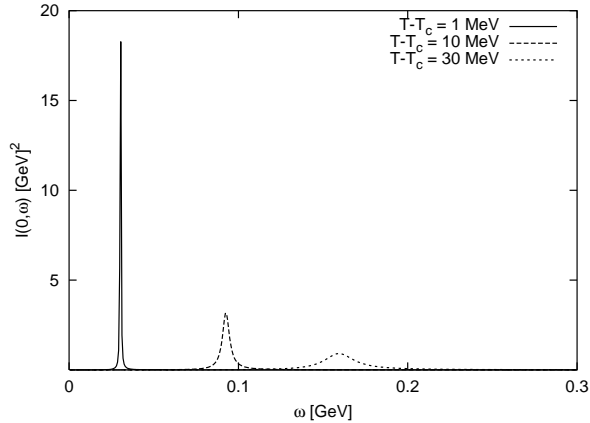}
\caption{The temperature dependence of the power spectrum at ${\bf k}={\bf 0}$ 
for a fixed chemical potentials $\mu = 280$ MeV.
The solid, dashed and dotted lines are for $T-T_c = 1$ MeV, $10$ MeV and $30$ MeV, 
respectively.}
\label{fig:PS_280}
\end{center}
\end{figure}

In order to clarify the equilibrium property above the critical temperatures, 
we calculated the temperature dependences of the power spectrum at 
$\mu = 280$ MeV in Fig.~\ref{fig:PS_280}.
Compared to Fig.~\ref{fig:PS_0}, 
one can see that 
above the second order phase transition, the temperature dependence of the 
power spectrum is almost same, although the peaks at $\mu = 280$ MeV 
approach to the origin faster.
However, interestingly enough, the behavior of the memory function shows that 
the relaxation is drastically changed around the tricritical point as 
we will see soon later.

We are interested in the behavior near the critical points and 
the typical time scale of the critical fluctuations increases 
because of the critical slowing down.
Then, we can ignore the microscopic time-dependence included 
in the memory function and the frequency function.
As is shown in Fig. \ref{fig:Renor}, the renormalized memory function 
converges to zero with a short time scale about 1 fm, 
that is negligible by comparison with the time scale of the critical fluctuations 
in the vicinity of the critical temperature.
Then, we can approximate the renormalized memory function as 
\begin{eqnarray}
\Phi({\bf k},t) \approx 2\gamma_{\bf k}\delta(t),
\end{eqnarray}
where
\begin{eqnarray}
\gamma_{\bf k} = \int^{\infty}_{0}d\tau \Phi({\bf k},\tau).
\end{eqnarray}
On the other hand, the frequency function converges to a finite value 
rapidly and we approximate it as 
\begin{eqnarray}
\Omega^2 ({\bf k},t) \approx \Omega^2_{\bf k} 
\equiv \lim_{t\rightarrow \infty}\Omega^2({\bf k},t).
\end{eqnarray}

These approximations are sometimes called the Markov limit.
Then, the Markovian Langevin equation is given by 
\begin{eqnarray}
\frac{d}{dt} \delta \sigma({\bf k},t) 
= - \Omega^2_{\bf k} \int^{t}_{0}d\tau \delta \sigma({\bf k},\tau) 
- \gamma_{\bf k} \delta \sigma({\bf k},t) 
+ \xi ({\bf k},t), \nonumber \\
 \label{eqn:MLLE} 
\end{eqnarray}
where
\begin{eqnarray}
\ll \xi({\bf k},t)\delta \sigma({\bf k}',0) \gg &=& \ll \xi({\bf k},t) \gg = 0, \\
\ll \xi({\bf k},t) \xi({\bf k}',t') \gg 
&=& 2V\gamma_{\bf k}\chi_{0}({\bf k})\delta^{(3)}_{\bf k,k'}\delta(t-t'). \nonumber \\
\end{eqnarray}
In the Markov limit, the colored noise is reduced to the Gaussian white noise.

There are two possibility of the solution of the Langevin equation, 
depending on the value of 
$A_{\bf k} = \sqrt{\Omega^2_{\bf k} - \gamma^2_{\bf k}/4}$.
The critical fluctuations relax exhibiting oscillation, if $A^2_{\bf k} \ge 0$.
However, if $A^2_{\bf k} \le 0$, this is the case of the overdamping and 
the solution does not show oscillation.
In our case, the condition $A^2_{\bf k} \ge 0$ is always satisfied.
Thus, the solution of this Langevin equation is 
\begin{eqnarray}
\lefteqn{ \ll \delta \sigma ({\bf k},t) \gg} && \nonumber \\
&=& \delta \sigma ({\bf k},0)  e^{-\gamma_{\bf k} t /2} \left[
-\frac{\gamma_{\bf k}}{2A_{\bf k}} 
\sin A_{\bf k}t
+ \cos A_{\bf k}t
\right].
\end{eqnarray}
From this analytic expression, one can see that the typical relaxation time 
of the long wave length component of the critical fluctuations 
is given by $\tau_{rt} = 2/\gamma_{\bf 0}$.

\begin{figure}\leavevmode
\begin{center}
\epsfxsize=9cm
\epsfbox{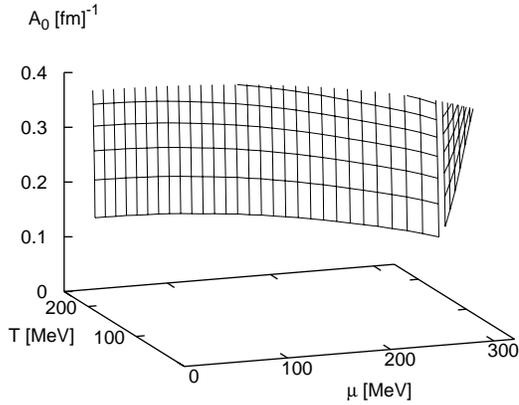}
\caption{Temperature and chemical potential dependences of 
the frequency $A_{\bf 0}$.}
\label{fig:Fre}
\end{center}
\end{figure}

First, we discuss the behavior of the frequency $A_{\bf k}$ 
that is shown in Fig.~\ref{fig:Fre} at a vanishing momentum.
Above the critical line of the second order phase transition, 
the frequency decreases at the almost same speed as the temperature 
approaches to the critical temperature for different $\mu$.
On the other hand, above the critical line of the first order phase transition, 
the frequency increases with $\mu$ along the critical line from the 
tricritical point.
This behavior is consistent with the behavior of the thermodynamic potential.
Above the second order phase transition, the thermodynamic potential has 
only one minimum and the curvature of the potential is negligibly small 
as is shown in Fig.~\ref{fig:EP}.
However, the thermodynamic potential has two minima above 
the first order phase transition and the curvature of the potential 
can be finite.

\begin{figure}\leavevmode
\begin{center}
\epsfxsize=9cm
\epsfbox{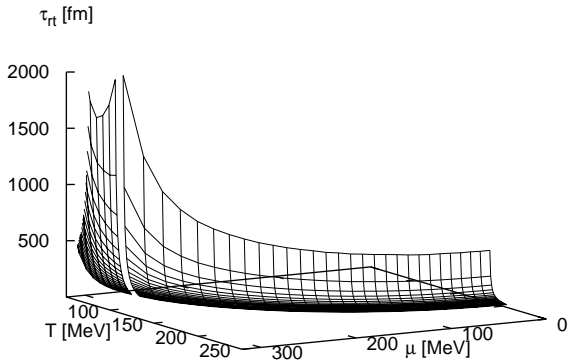}
\caption{Temperature and chemical potential dependences of the 
relaxation time at higher temperature than 50 MeV.}
\label{fig:LT}
\end{center}
\end{figure}

\begin{figure}\leavevmode
\begin{center}
\epsfxsize=9cm
\epsfbox{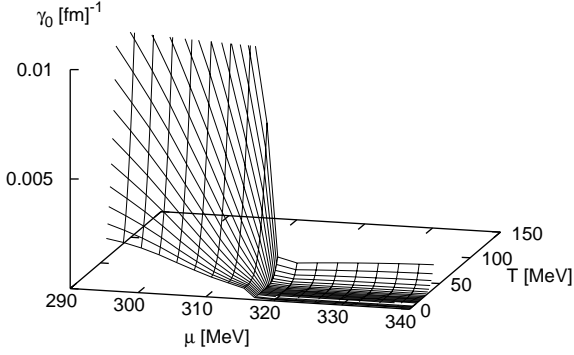}
\caption{The temperature and chemical potential dependences of the 
damping coefficient $\gamma_{\bf 0}$ in low temperature and 
large chemical potential region.}
\label{fig:Damp}
\end{center}
\end{figure}

The behavior of the relaxation is extremely interesting 
in comparison with that of the frequency.
The temperature and chemical potential dependences of the relaxation time 
is shown in Fig.~\ref{fig:LT}.
The relaxation time at lower temperature than 50 MeV is not shown 
because it diverges at speed.
Thus, we plotted the damping coefficient $\gamma_{\bf 0}$ in low temperature 
and large chemical potential region 
in Fig.~\ref{fig:Damp} instead of the relaxation time.
At a fixed $\mu$, the relaxation time increases as the temperature 
is lowered toward the critical temperature and diverges at the 
critical point because of the critical slowing down.
In this figure, we started to plot from the 1 MeV higher temperature than 
the critical temperature.
When we increase $\mu$ along the critical line, 
we encounter the enhancement of the critical slowing down;
one is in the low temperature and large chemical potential region and 
the other is around the tricritical point.
The former can be explained by the Pauli blocking due to 
the Fermi surface.
As a matter of fact, 
the damping coefficient vanishes quickly there as is shown in Fig.~\ref{fig:Damp}.

More interestingly, the critical slowing down can be enhanced 
around the tricritical point.
This would be related to the broadness of the bottom of the 
thermodynamic potential.
As is shown in Fig. \ref{fig:EP}, 
when we increase the chemical potential, the bottom of the 
thermodynamic potential increases gradually and achieves maximum 
at the tricritical point.
At larger chemical potential than that of the tricritical point, 
the thermodynamic potential has two minimum but the potential 
barrier between the two minima is still shallow.
Thus, the large fluctuations are still possible to survive.
In short, there exist extraordinary large fluctuations around the tricritical point 
and this causes the enhancement.
Because of the large relaxation time of the critical fluctuations, 
thermalization is decelerated, in particular, 
in the low temperature and large chemical potential region and around 
the tricritical point.

It should be noted that the enhancement is not obvious 
in the behavior of the power spectrum.
This is because the power spectrum is given by $\Phi({\bf k},\omega)$ 
and $\Omega^2_{\bf k}(\omega)$, and the former is 
smaller than the latter.
Thus, the behavior of the power spectrum is dominated by $\Omega^2_{\bf k}(\omega)$ 
and the information of $\Phi({\bf k},\omega)$ is smeared.
However, as is indirectly shown in Fig. \ref{fig:Fre}, 
the frequency $\Omega^2_{\bf k}(\omega)$ does not reveal the anomalous 
behavior around the tricritical point, clearly.
Thus, we cannot see the remarkable difference between Figs. \ref{fig:PS_0} 
and \ref{fig:PS_280}.

\section{Concluding remarks} \label{chap:Sum}

We have derived the linear Langevin equation 
that describes the dynamics of the chiral phase transition 
at finite temperature and density.
The simple application of the projection operator method 
to the Nambu-Jona-Lasinio (NJL) model 
caused the long time correlation in the memory function.
To avoid this difficulty, we introduced the renormalization of the 
memory function.
The resulting Langevin equation reveals 
the critical slowing down at the same temperature 
as that estimated in the mean-field approximation and shows thermalization.

The order parameter relaxes exhibiting oscillation and 
this is different from the behavior expected from 
a simple diffusion-type equation like the time-dependent Ginzburg-Landau 
(TDGL) equation .
This means that the assumption employed to derive the TDGL equation 
may lose its validity in the chiral phase transition.
In the derivation of the TDGL equation, one assumes that 
the time evolution of the order parameter is induced by 
a sort of the thermodynamic force that is given by $\delta F/\delta M$, 
where $F$ is the Ginzburg-Landau free energy and $M$ is the order parameter.
However, in general, a number of thermodynamic forces are possible 
to induce such an irreversible process 
because of the cross effect ( 
for example, the Dufour effect and the Serot effect in 
thermal conduction).
Then, the usual TDGL equation can be changed from a simple diffusion-type 
equation.

The power spectrum was also calculated 
in different two ways.
One is the application of the Wiener-Khinchin theorem and 
the other is to solve the evolution equation of the correlation function 
under the assumption of initial thermal equilibrium.
The two power spectrums have completely same form and 
we can conclude that the Langevin equation shows thermalization.
The energy and temperature dependences of the power spectrum 
is consistent with that calculated in the linear response theory 
where the random phase approximation is employed
\cite{ref:HK} 
and there exists the soft mode. 
As a result, we can conclude that our Langevin equation 
fulfils the requirements near $T_{c}$.

In this calculation, 
we approximate the memory function calculating the ring diagram contribution 
because we want to derive the equation that converges to 
the equilibrium state consistent with the mean-field approximation.
However, it is possible to derive the equation including 
higher order contributions.
Then, the merit of our formulation is to be possible 
to derive {\bf nonperturbative} results by approximating the memory function 
{\bf perturbatively}, 
and the higher order term is already known by the systematic expansion
of ${\mathcal D}(t,t_0)$.

In the latter part of this paper, 
we discussed the temperature and chemical potential dependences 
of the critical fluctuations in the chiral phase transition.
At a fixed $\mu$, the relaxation time of the critical fluctuations 
increases because of the critical slowing down.
On the other hand, 
increasing the chemical potential along the critical line, 
we found the enhancement of the critical slowing down.
One is seen in low temperature and large chemical potential 
region.
This is because the decay of the critical fluctuations is suppressed 
owing to the Pauli blocking.
The other enhancement is shown around the tricritical point.
This would be related to the broadness of the thermodynamic potential 
around the tricritical point and 
means that there exist extremely large fluctuations.
The enhancement of the critical slowing down 
may affect the formation of the disoriented chiral condensate (DCC).
Around the tricritical point, 
the large fluctuations can have a long relaxation time and 
it may be of advantage to form the large domain of DCC.

As is well-known, the mean-field approximation cannot describe the static and 
dynamical critical exponents correctly in various cases.
Thus, the validity of our results might be suspected.
However, it should be noted that we approximate the memory function by 
the ring diagram contribution and hence the relaxation dynamics itself 
includes the fluctuation effects, although the equilibrium state is 
consistent with the mean-field one.

As another problem, 
we used the NJL model as a low-energy effective theory of QCD, 
where the endpoint is located at lower temperature 
and higher chemical potential than those estimated in other approaches, 
for example, Lattice QCD calculations
\cite{ref:Fodor,ref:Fodor2,ref:Karsch,ref:de,ref:Elia,ref:de2,ref:Fodor3}.
However, we can expect that our calculation is still available to 
describe the critical dynamics qualitatively, 
because the enhancement of the critical slowing down can be explained 
from the Pauli blocking and the behavior of the thermodynamic potential.
Thus, it is reasonable to expect 
that the behavior still survive even if we take into account 
higher order corrections or calculate in more realistic models.

In this study, we ignored the finite current quark mass.
When we take into account it, 
the phase transition is changed from the second order to crossover 
and the tricritical point is replaced by the endpoint.
Then, the critical slowing down is smeared and hence the 
behavior around the endpoint will be changed.

We have discussed the derivation and the behavior of 
the Langevin equation ignoring nonlinear effects.
However, 
from the Ginzburg criterion, 
the nonlinear fluctuations play important roles 
near the critical points.
As a matter of fact, 
the mode coupling theory makes it clear that to calculate the 
dynamical critical exponent correctly, we must introduce 
nonlinear terms in the van Hove theory
\cite{ref:Kawasaki}.
The nonlinear term will affect not only the dynamical critical exponent 
but also the power spectrum.
The behavior of the power spectrum discussed in this paper 
was almost same as that in the linear response theory.
This means that the conventional discussions based on the linear response theory 
may be inadequate to discuss the behaviors of the critical fluctuations.
Furthermore, by using the {\it nonlinear} Langevin equation, 
we can discuss the possibility of the dynamical transition in QCD 
like the ergodic-nonergodic transition 
that is discussed in the glass transition
\cite{ref:glass}.
In the projection operator method, 
the effect of the nonlinearity can be incorporated by 
introducing nonlinear terms as gross variables in the definition of 
the Mori projection operator.

\begin{figure}\leavevmode
\begin{center}
\epsfxsize=5cm
\epsfbox{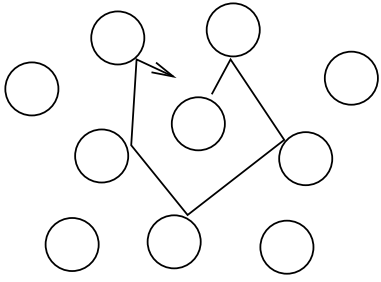}
\caption{
A particle can exchange its energy and momentum 
by collisions, but it is difficult to move far away from the initial position 
in high density system.}
\label{fig:Jam}
\end{center}
\end{figure}

In this paper, we have assumed that there is only one gross variable, that is, 
the fluctuations of the order parameter.
However, densities of conserved quantities also can be gross variables 
as is discussed in Sec. \ref{chap:Mori}.
The most promising candidate is the number density of quarks.
In the high density system, 
a particle is thickly surrounded by other particles.
The energy and momentum of a particle is exchanged continuously by 
collisions, but it is difficult to move far away from the initial position, 
as is schematically shown in Fig. \ref{fig:Jam}.
Therefore, the number density changes slowly and can be a gross variable 
compared to the energy and momentum densities.
This situation is probably realized in glass transition 
where the fluctuations of the number density play a role of 
an order parameter
\cite{ref:glass}.
As a matter of fact, 
the effect of the number density for the critical fluctuations 
has been studied in the linear response theory
\cite{ref:Fujii,ref:Fujii2}.

Definitely, our discussion is applicable to other phase transition 
like the color superconducting phase transition.
These subjects are future problems.

\vspace*{1cm}

T.K. thanks A.~Muronga and 
D.~Rischke for fruitful discussions and comments.
T.K. acknowledges a fellowship from the Alexander von Humboldt 
Foundation.

\appendix

\section{Fluctuation-dissipation theorem of second kind}\label{app:FD}

Sometimes, a Langevin equation is phenomenologically derived, where 
a noise term is artificially introduced to realize thermalization.
Then, we assume that there exists a relation between a memory function and 
a noise term like the Einstein relation in the Brownian motion.
However, in this projection operator method, 
we can prove the exact relation between a noise term and a memory term.
This is called the fluctuation-dissipation theorem of second kind
\cite{ref:Mori}.
In the following discussion, we use the Mori projection operator 
that is defined in Sec. \ref{chap:Mori}.

Let us regard $A({\bf x})$ as a gross variable and derive a Langevin equation 
for $A({\bf x})$.
For simplicity, we assume that $A({\bf x})$ does not have a finite value 
in equilibrium, $\langle A({\bf x}) \rangle_{eq} = 0$.
Then, the Mori projection is defined by 
\begin{eqnarray}
P~O = (O,A({\bf x}'))\cdot (A({\bf x}'),A({\bf x}''))^{-1}\cdot A({\bf x}'').
\end{eqnarray}
Substituting this expression into the exact TC equation 
(\ref{eqn:TC-1}), we have 
\begin{eqnarray}
\lefteqn{ \frac{d}{dt}A({\bf x},t) } && \nonumber \\
&=& (iLA({\bf x}),A^{\dagger}({\bf x}'))\cdot 
(A({\bf x}'),A^{\dagger}({\bf x}''))\cdot A({\bf x}'',t) \nonumber \\
&&+ \int^{t}_{t_{0}}ds (iLf({\bf x},s),A^{\dagger}({\bf x}'))\cdot 
(A({\bf x}'),A^{\dagger}({\bf x}''))^{-1}\cdot A({\bf x}'',t-s) 
+ f({\bf x},t) \nonumber \\
&=& (iLA({\bf x}),A^{\dagger}({\bf x}'))\cdot 
(A({\bf x}'),A^{\dagger}({\bf x}''))\cdot A({\bf x}'',t) \nonumber \\
&&+ \int^{t}_{t_{0}}ds (f({\bf x},s),f^{\dagger}({\bf x}',0))\cdot 
(A({\bf x}'),A^{\dagger}({\bf x}''))^{-1}\cdot A({\bf x}'',t-s) 
+ f({\bf x},t). \nonumber \\
\end{eqnarray}
Here, we put $A({\bf x}) = A({\bf x},t_{0})$, for simplicity.
The operator form of the noise is given by
\begin{eqnarray}
f({\bf x},t) = Qe^{iLQ(t-t_{0})}iLA({\bf x}),
\end{eqnarray}
that has the following properties:
\begin{eqnarray}
(f({\bf x},t),A({\bf x}')) &=& 0, \\
\langle f({\bf x},t) \rangle_{eq} 
&\equiv& {\rm Tr}[\rho f({\bf x},t)] = 0,
\end{eqnarray}
where $\rho$ is given by $\rho = e^{-\beta H}/{\rm Tr}[e^{-\beta H}]$.
Here we used that $PQ =0$.
The first correlation means that the noise term is always orthogonal to the 
gross variable $A({\bf x})$.
The second correlation indicates that the thermal average of the noise vanishes.

Now, we introduce the memory function as
\begin{eqnarray}
\Gamma({\bf x},{\bf x}'';t,0)
= (f({\bf x},t),f^{\dagger}({\bf x}',0))\cdot 
(A({\bf x}'),A^{\dagger}({\bf x}''))^{-1}. \nonumber \\
\label{eqn:2FDT}
\end{eqnarray}
This exact Langevin equation with the Mori projection is, sometimes, called 
the Mori equation.
We can see that the memory function is given by the time correlation of the noise 
in the Mori equation.
This exact relation is called the fluctuation-dissipation theorem of 
second kind (2nd F-D theorem).
In this paper, we assume that the exact relation should be satisfied 
even for the approximated memory function.
Then, the noise is determined so as to reproduce the 2nd F-D theorem, that is, 
condition (\ref{eqn:2FDT}).

\section{Derivation of Mori projection operator} \label{app:MPRP}

We shall consider the system which is in the thermal equilibrium state 
with small external perturbation $\{ h_{\mu} \}$ at $t < 0$.
At $t=0$, the external perturbation is switched off, and 
the system relaxes to a new thermal equilibrium state at $t > 0$.
The initial density matrix to describe this situation is 
\begin{eqnarray}
\rho(\{ h_{\mu} \}) = \frac{1}{Z_{h}} 
\exp [-\beta(H-\sum_{\mu}\int d^3 {\bf x}h_{\mu}({\bf x})A_{\mu}({\bf x}))],
\nonumber \\
\end{eqnarray}
where $\{ A_{\mu}({\bf x}) \}$ is a dynamical variable, 
whose relaxation we are interested in and $Z_{h}$ is the normalization factor.

The most probable relaxation path at $t > 0$ in the Heisenberg picture, 
is given by
\begin{eqnarray}
\langle A_{\mu}({\bf x},t) \rangle 
&=& \Tr [\rho(\{ h_{\mu} \}) A_{\mu}({\bf x},t)]. \label{eqn:MPRP}
\end{eqnarray}
At $t > 0$, the dynamics of the system is governed by the Hamiltonian $H$.
The time evolution of the dynamical variable, thus, is defined by 
\begin{eqnarray}
A_{\mu}({\bf x},t) = e^{iHt}A_{\mu}({\bf x})e^{-iHt},
\end{eqnarray}
where $A_{\mu}({\bf x}) = A_{\mu}({\bf x},0)$.
When the initial deviation from thermal equilibrium is small, 
the external perturbation $\{h_{\mu}\}$ is small and 
Eq. (\ref{eqn:MPRP}) can be approximated as
\begin{eqnarray}
\langle A_{\mu}({\bf x},t) \rangle 
&=& 
\langle A_{\mu}({\bf x}) \rangle_{0} \nonumber \\
&&\hspace*{-1.5cm} + \sum_{\nu}\int d^3 {\bf x'}
\beta( A_{\mu}({\bf x},t), 
~A_{\nu}({\bf x'})-\langle A_{\nu}({\bf x'}) \rangle_{0} )_{0}
h_{\nu}({\bf x'}), \nonumber \\
\label{eqn:MPRP-1st}
\end{eqnarray}
where $\langle O \rangle_{0}$ means $\Tr [\rho(0) O]$ and the 
inner product is the Kubo's canonical correlation defined in Eq. (\ref{eqn:KCC});
\begin{eqnarray}
(F,~G)_{0} = \int^{\beta}_{0}\frac{d\lambda}{\beta}
\Tr [\rho(0) e^{-\lambda H} F e^{\lambda H} G].
\end{eqnarray}
The small external perturbation $\{ h_{\mu} \}$ 
in Eq. (\ref{eqn:MPRP-1st}) is eliminated 
using Eq. (\ref{eqn:MPRP-1st}) at $t=0$.
Thus, we have
\begin{eqnarray}
\langle A_{\mu}({\bf x},t) \rangle 
= \sum_{\mu,\nu} \int d^3 {\bf x'} d^3 {\bf x''}
( A_{\mu}({\bf x},t),~A_{\gamma}({\bf x'}) )
\cdot [(A({\bf x'}),~A({\bf x''}))^{-1}]_{\gamma,\nu}
\cdot \langle A_{\mu}({\bf x},0) \rangle, \nonumber \\
\label{eqn:MPP}
\end{eqnarray}
where the inverse is defined by
\begin{eqnarray}
\sum_{\nu}\int d^3 {\bf x}
[(A({\bf x}),A({\bf x'}))^{-1}]_{\mu,\nu} \cdot
(A_{\nu}({\bf x'}),A_{\gamma}({\bf x''})) 
= \delta_{\mu,\gamma}\delta^{(3)}({\bf x}-{\bf x''}).
\end{eqnarray}
Here, we set $\langle A_{\mu}({\bf x}) \rangle_{0}=0$, for simplicity.

This equation (\ref{eqn:MPP}) can be reexpressed as $PA_{\mu}({\bf x},t)$ 
by introducing the Mori projection operator,
\begin{eqnarray}
P O &=& \sum_{\mu,\nu} \int d^3 {\bf x'} d^3 {\bf x''}
( O,~A_{\gamma}({\bf x'}) )
\cdot [(A({\bf x'}),~A({\bf x''}))]^{-1}_{\gamma,\nu} \cdot A_{\mu}({\bf x}). 
\nonumber \\
\end{eqnarray}
Namely, the relaxation of an arbitrary operator is extracted by 
operating the Mori projection operator.

\section{The streaming term}\label{app:stream}

First, it should be noted that the order parameter commutes with the 
interaction part;
\begin{eqnarray}
iL_{I}\delta \sigma({\bf x}) = 0.
\end{eqnarray}
Thus, the streaming term is given by
\begin{eqnarray}
e^{iLt}PiL_{0}\delta \sigma({\bf x}) 
&=& e^{iLt}\lim_{t \rightarrow 0}P\frac{d}{dt}\delta \sigma_{0}({\bf x},t) \nonumber \\
&=& e^{iLt}\lim_{t \rightarrow 0} 
\int d^3{\bf x}'d^3 {\bf x}''\frac{d}{dt}\chi_{t}({\bf x-x'})
\chi^{-1}_{0}({\bf x'-x''}) \delta \sigma({\bf x}''), \nonumber \\
\label{eqn:Stream1}
\end{eqnarray}
where 
\begin{eqnarray}
\chi_{t-t'}({\bf x-x'})
&=& (\delta \sigma_0 ({\bf x},t),\delta \sigma_0 ({\bf x}',t')) \nonumber \\
&=& 
\int^{\beta}_{0}\frac{d\lambda}{\beta}
\{
\contraddd{\bar{q}^\alpha_{0}({\bf x},t-i\lambda)}{q^\beta_{0}({\bf x}',t')}
\hspace{1.4cm}
\contraddd{q^\alpha_{0}({\bf x},t-i\lambda)}{\bar{q}^\beta_{0}({\bf x}',t')}
\hspace{1.4cm}
\} \nonumber \\
&=& 
\frac{N_c N_f}{\beta V^2}\sum_{\bf p,p'}\{
(\sin^2 \theta^+_{p'} - \sin^2 \theta^+_{p})
\frac{p\cdot p' + M^2}{E_p E_p' (E_{p}-E_{p'})}e^{i(E_{p}-E_{p'})(t-t')}
e^{-i{(p-p')(x-x')}} \nonumber \\
&& + (\sin^2 \theta^-_{p} - \sin^2 \theta^-_{p'})
\frac{-p\cdot p' - M^2}{E_p E_p' (E_{p}-E_{p'})}e^{-i(E_{p}-E_{p'})(t-t')}
e^{i{(p-p')(x-x')}} \nonumber \\
&& + (1 - \sin^2 \theta^+_{p} - \sin^2 \theta^-_{p'})
\frac{p\cdot p' - M^2}{E_{p}E_{p'}(E_{p}+E_{p'})} \nonumber \\
&& \times(e^{i(E_{p}+E_{p'})(t-t')} + e^{-i(E_{p}+E_{p'})(t-t')})e^{-i{\bf (p+p')(x-x')}}
\}, \nonumber \\
\end{eqnarray}
and
\begin{eqnarray}
\int d^3 {\bf x}' \chi_{0}({\bf x-x'})\chi^{-1}_{0}({\bf x'-x''}) 
&=& \delta^{(3)}({\bf x-x''}).
\end{eqnarray}
Here, we used the contraction defined in Appendix \ref{app:Propa}.
Substituting the above expression into Eq. (\ref{eqn:Stream1}),
it is easy to show that
\begin{eqnarray}
\lim_{t\rightarrow 0}\frac{d}{dt}\chi_{t}({\bf x-x'}) = 0.
\end{eqnarray}
Thus, the contribution from the streaming term vanishes.

\section{Contraction of fermion field at finite temperature and density}\label{app:Propa}

In this Appendix, we introduce the contraction of the fermion field.
This is convenient to implement the calculation 
with projection operator and 
we can regard it as a correspondence 
of the propagator in the usual finite temperature field theory
\cite{ref:Koide}.

First, we introduce the Bogoliubov transformation through which 
the new pairs of creation and annihilation operators are defined
\cite{ref:TFD}:
\begin{eqnarray}
b_{{\bf k},s} &=& \cos \theta^+_{k}B_{{\bf k },s} + 
i\sin \theta^+_{k} \tilde{B}^{\dagger}_{{\bf k },s}, \\
d_{{\bf k},s} &=& \cos \theta^-_k D_{{\bf k },s} + 
i\sin \theta^-_k \tilde{D}^{\dagger}_{{\bf k },s},
\end{eqnarray}
where
\begin{eqnarray}
\sin^2 \theta^{\pm}_{\bf k} = (e^{\beta \epsilon^{\mp}_{k}} + 1)^{-1} 
\equiv n^{\pm}(E_{k}), 
\end{eqnarray}
and 
\begin{eqnarray}
\epsilon^{\pm}_{k} = \sqrt{k^2 + M^2} \pm \mu \equiv E_{k} \pm \mu.
\end{eqnarray}
Here, $M$ is the mass of the fermion.
The above operators satisfy the commutation relation
\begin{eqnarray}
[B_{{\bf k},s},B^{\dagger}_{{\bf k}',s'}]_+
&=& [\tilde{B}_{{\bf k},s},\tilde{B}^{\dagger}_{{\bf k}',s'}]_+ 
= \delta_{s,s'}\delta^{(3)}_{\bf k,k'}, \\
\protect{[D_{{\bf k},s},D^{\dagger}_{{\bf k}',s'}]_+} 
&=& [\tilde{D}_{{\bf k},s},\tilde{D}^{\dagger}_{{\bf k}',s'}]_+ 
= \delta_{s,s'}\delta^{(3)}_{\bf k,k'}.
\end{eqnarray}
All the other commutators vanish.
Now, we can define the thermal vacuum $|\theta \rangle$ as
\begin{eqnarray}
B_{{\bf k},s}|\theta \rangle
= \tilde{B}_{{\bf k},s}|\theta \rangle
= \protect{D_{{\bf k},s}|\theta \rangle}
= \protect{\tilde{D}_{{\bf k},s}|\theta \rangle}
=0.
\end{eqnarray}
Then, we can express the statistical average in terms 
of the vacuum expectation value:
\begin{eqnarray}
{\rm Tr}[\rho_{eq} O] = \langle \theta |O |\theta \rangle,
\end{eqnarray}
where $\rho_{eq}$ is a density matrix of a thermal equilibrium state.
Thus, we can use Wick's theorem that simplifies 
the calculation of the correlation function.
Wick's theorem tells us that arbitrary correlation functions are 
expressed as the sum of all possible products of contractions.
In our calculation, contractions are given by
\begin{eqnarray}
\lefteqn{\contra{q_{0}({\bf x},t)}{\bar{q}_{0}({\bf x'},t')}} && \nonumber \\
&=& \frac{1}{V}\sum_{\bf p}\{ \cos^2 \theta^+_{p} \frac{p\hspace{-0.15cm}/ + M}{2E_{p}}
e^{i{\bf p(x-x')}} e^{-i\epsilon^-_{p}(t-t')}
+ \sin^2 \theta^-_{p}\frac{p\hspace{-0.15cm}/ - M}{2E_{p}}
e^{-i{\bf p(x-x')}} e^{i\epsilon^+_{p}(t-t')}\}, \\
\lefteqn{\contrad{\bar{q}_{0}({\bf x}',t')}{q_{0}({\bf x},t)}} && \nonumber \\
&=& \frac{1}{V}\sum_{\bf p}\{ \sin^2 \theta^+_{p} 
\left(\frac{p\hspace{-0.15cm}/ + M}{2E_{p}}\right)^T
e^{i{\bf p(x-x')}} e^{-i\epsilon^-_{p}(t-t')}
+ \cos^2 \theta^-_{p} \left(\frac{p\hspace{-0.15cm}/ - M}{2E_{p}}\right)^T
e^{-i{\bf p(x-x')}} e^{i\epsilon^+_{p}(t-t')}\}. \nonumber \\
\end{eqnarray}

\section{Brownian motion in harmonic potential}\label{app:Brown}

As an exact solvable example, 
let us consider the one-dimensional classical system where 
a heavy particle with the mass $M$ interacts with heat bath composed of 
many harmonic oscillators
\cite{ref:Zwanzig}.
The Hamiltonian of the total system is 
\begin{eqnarray}
H = \frac{p^2}{2M} + \frac{M\omega_{0}^2}{2}x^2 
+ \sum_{i}[
\frac{p_{i}^2}{2} + \frac{\omega_{i}^2}{2}
(x_{i}-\frac{\gamma_{i}}{\omega_{i}^2}x)^2
].
\end{eqnarray}
The Newton equations of this system are given by 
\begin{eqnarray}
&&\dot{x} = \frac{p}{M},~~~~~\dot{p} = -M\omega^2_{0}x 
+ \sum_{i}\gamma_{i}(x_{i}-\frac{\gamma_{i}}{\omega_{i}^2}x), \\
&&\dot{x}_{j} = p_{j},~~~~~~~\dot{p}_{j} = 
-\omega^2_{j}(x_{j}-\frac{\gamma_{j}}{\omega^2_{j}}x).
\end{eqnarray}

The mass of the particle $M$ is large and the coordinate $x$ and 
the momentum $p$ change slowly in comparison with $p_i$ and $x_i$.
In this case, it is natural to choose $p$ and $x$ 
as gross variables of this system.
Then, the Mori projection operator is defined by 
\begin{eqnarray}
P G 
&=&  (G,x)(x,x)^{-1}x + (G,p)(p,p)^{-1}p \nonumber \\
&=&  \beta M\omega^2_0 x (G,x) + \frac{p\beta}{M} (G,p).
\end{eqnarray}
The inner product is given by the canonical correlation for the classical variables,
\begin{eqnarray}
(X,Y) = \int^{\beta}_{0}\frac{d\lambda}{\beta} 
{\rm Tr}[\rho e^{\lambda H}Xe^{-\lambda H}Y ] 
= {\rm Tr}[\rho XY] 
\equiv \langle XY \rangle,
\end{eqnarray}
where $\rho = e^{-\beta H}/{\rm Tr}[e^{-\beta H}]$.
Then, we have the following correlations:
\begin{eqnarray}
( p,p ) &=& M/\beta, \\
( \dot{p},x ) &=& -( p,\dot{x} ) = -( p,p )/M, \\
( (x_{i}-\gamma_{i}x/\omega_{i}^2),x ) &=& ( p,x ) = 0, \\
( x,x ) &=& (M\omega^2_{0}\beta)^{-1}.
\end{eqnarray}

When we substitute the Mori projection operator into the TC equation 
and set $O(0) = x(0)$ and $p(0)$, 
we have the following Langevin equation 
\begin{eqnarray}
\frac{d}{dt} x(t) &=& \frac{p(t)}{M}, \\
\frac{d}{dt} p(t) &=& -M\omega^2_0 x(t) - \int^{t}_0 ds \Xi(t-s) p(s) + f(t), 
\end{eqnarray}
where
\begin{eqnarray}
\Xi(t) 
&=& \frac{1}{M}\sum^{N}_{j=1}\frac{\gamma_i}{\omega^2_j}\cos \omega_j t, \\
f(t)
&=& \sum^{N}_{j=1}\gamma_i[ \cos \omega_j t 
(x_i(0) - \frac{\gamma_j}{\omega^2_j}x(0)) + \frac{\sin \omega_j t}{\omega_j}p_j (0) ]. 
\end{eqnarray}

We assume that the frequency of the oscillations distribute continuously 
\begin{eqnarray}
g(\omega) = \left\{
\begin{array}{cc}
\frac{\omega^2}{\omega^3_d}, & \omega < \omega_d \\
0 & \omega > \omega_d ,
\end{array}
\right.
\end{eqnarray}
where $\omega_d$ is a cutoff.
Then, the memory function is given by
\begin{eqnarray}
\Xi (t) 
&=& \frac{1}{M}\int^{\omega_d}_{0} d\omega 
\frac{\gamma}{\omega^3_d}\cos \omega t
\nonumber \\
&=& 2\Xi_0 \frac{\sin \omega_d t}{\pi t},
\end{eqnarray}
where $\Xi_0 = \frac{\pi \gamma^2}{2 \omega^3_d M}$.

When we are interested in the motion with a macroscopic time scale 
larger than $2\pi/\omega_d$, the memory function is approximated by
\begin{eqnarray}
\Xi(t) \approx 2\Xi_0 \delta (t).
\end{eqnarray}
Finally, the Langevin equation is give by 
\begin{eqnarray}
\frac{d}{dt} x(t) &=& \frac{p(t)}{M}, \\
\frac{d}{dt} p(t) &=& -M\omega^2_0 x(t) - \Xi_0 p(t) + f(t),
\end{eqnarray}
where the noise has the following correlation properties;
\begin{eqnarray}
\langle f(t) \rangle &=& 0, \\
\langle f(t)f(t') \rangle &=& \frac{2M}{\beta}\Xi_0 \delta (t-t').
\end{eqnarray}

In order to see that the behavior of the memory function strongly depends on 
the choice of the projection operator, 
we derive the Langevin equation with the incomplete Mori projection 
defined by using one gross variable,
\begin{eqnarray}
P G 
=  (G,p)(p,p)^{-1}p 
=  \frac{p\beta}{M} (G,p).
\end{eqnarray}
Then, we have the Langevin equation with one gross variable $p$,
\begin{eqnarray}
\frac{d}{dt}p(t) = -\int^{t}_{0}ds \Xi'(t-s) p(s) + f'(t),
\end{eqnarray}
where
\begin{eqnarray}
\Xi'(t) &=& \omega^2_0 + \Xi(t), \\
f'(t) &=& -M \omega^2_0 x(0) + f(t).
\end{eqnarray}
It is clear that $\Xi'(t)$ converges to $\omega^2_0$ 
at $t \rightarrow \infty$, although $\Xi(t)$ vanishes.
Namely, the behavior of the memory function depends on the definition of 
the Mori projection and we must choose the complete set 
of the gross variables to derive the memory function without 
the long time correlation.


\begin{thebibliography}{99}
%
\bibitem{ref:Bearden}
I.~G.~Bearden et al., 
Phys.~Rev.~Lett.~{\bf 78} (1997) 2080.
%
\bibitem{ref:Roland}
H.~Appelsh\"auser et al., 
Nucl.~Phys.~{\bf A638} (1998) 91c.
%
\bibitem{ref:Kolb}
P.~F.~Kolb, P.~Huovinen, U.~Heinz and H.~Heiselberg,
Phys.~Lett.~{\bf B500} (2001) 232.
%
\bibitem{ref:Raja}
M.~ A.~ Stephanov, K.~ Rajagopal and E.~ V.~ Shuryak,
Phys.~Rev.~Lett.~{\bf 81} (1998) 4816.
%
\bibitem{ref:Raja2}
M.~ A.~ Stephanov, K.~ Rajagopal and E.~ V.~ Shuryak,
Phys.~Rev.~{\bf D60} (1999) 114028.
%
\bibitem{ref:SS1}
D.~T.~Son and M.~A.~Stephanov, Phys.~Rev.~Lett. {\bf 88} (2002) 202302.
%
\bibitem{ref:SS2}
D.~T.~Son and M.~A.~Stephanov, Phys.~Rev. {\bf D66} (2002) 076011.
%
\bibitem{ref:Rischke}
O.~ Scavenius, A.~ Mocsy, I.~N.~ Mishustin and D.~H.~ Rischke,
Phys.~Rev.~{\bf C64} (2001) 045202.
%
\bibitem{ref:Hatta}
Y.~Hatta and T.~Ikeda, Phys.~Rev.~{\bf D67} (2003) 014028.
%
\bibitem{ref:Fujii}
H.~Fujii, Phys.~Rev.~{\bf D67} (2003) 094018.
%
\bibitem{ref:Fujii2}
H.~Fujii and M.~Ohtani, hep-th/0402263.
%
\bibitem{ref:Paech}
K.~Paech, H.~St\"ocker and A.~Dumitru, Phys.~Rev.~{\bf C68} (2003) 044907.
%
\bibitem{ref:Kodama}
C.~E.~Aguiar, E.~S.~Fraga and T.~Kodama, nucl-th/0306041.
%
\bibitem{ref:Boya0}
D.~Boyanovsky, H.~J.~de Vega, R.~Holman and M.~Siminato, 
Phys.~Rev.~{\bf D60} (1999) 065003.
%
\bibitem{ref:Boya1}
D.~Boyanovsky, H.~J.~de Vega and M.~Siminato, Phys.~Rev.~{\bf D63} (2001) 045007.
%
\bibitem{ref:Boya2}
D.~Boyanovsky and H.~J.~de Vega, Phys.~Rev.~{\bf D65} (2002) 085038.
%
\bibitem{ref:Boya3}
D.~Boyanovsky and H.~J.~de Vega, Ann.~Phys.~(N.Y.) {\bf 305} (2003) 335.
%
\bibitem{ref:Boya}
D.~Boyanovsky, H.~J.~de Vega and S.~Y.~Wang, hep-ph/0312185.
%
\bibitem{ref:Kawasaki}
K.~Kawasaki, Ann.~Phys.~{\bf 61} (1970) 1.
%
\bibitem{ref:Kawasaki2}
K.~Kawasaki, in {\it Phase transition and Critical Phenomena}, edited by 
C.~Domb and M.~S.~Green (Academic press, London, 1976).
%
\bibitem{ref:Kawasaki3}
K.~Kawasaki and 
J.~D.~Gunton, in {\it Progress in Liquid Physics}, edited by C.~A.~Croxton
(Wiley, New York, 1978).
%
\bibitem{ref:Muronga}
As for the effect of dissipation in the relativistic system, 
see, A.~Muronga, nucl-th/0309055.
%
\bibitem{ref:Cooper}
L.~M.~A.~Bettencourt, F.~Cooper and K. Pao,
Phys.~Rev.~Lett. {\bf 89} (2002) 112301.
%
\bibitem{ref:HH}
P.~C.~Hohenberg and B.~I.~Halperin, Rev.~Mod.~Phys. {\bf 49} (1977) 435.
%
\bibitem{ref:Onuki}
A.~Onuki, {\it Phase Transition Dynamics}, (Cambridge, UK, 2002)
%
\bibitem{ref:Wakou}
There is another method to derive the Langevin equation.
See, J.~Wakou, J,~Koide and R.~Fukuda, Physica {\bf A276} (2000) 164.
%
\bibitem{ref:Morikawa} 
M.~Morikawa,~Phys.~Rev.~{\bf D33} (1986) 3607.
%
\bibitem{ref:Glei-Ramo} 
M.~Gleiser and R.~O.~Ramos,~Phys.~Rev.~{\bf D50} (1994) 2441.
%
\bibitem{ref:Ber-Glei-Ramo} 
A.~Berera, M.~Gleiser and R.~O.~Ramos,~Phys.~Rev.~{\bf D58} (1998) 123508.
%
\bibitem{ref:Lom-Mazz}  
F.~Lombardo and F.~D.~Mazzitelli,~Phys.~Rev.~{\bf D53} (1996) 2001.
%
\bibitem{ref:Grei-Mul}  
C.~Greiner and B.~M\"uller,~Phys.~Rev~{\bf D55} (1997) 1026.
%
\bibitem{ref:Grei-Leu}
C.~Greiner and S.~Leupold, Ann.~Phys.~{\bf 270} (1998) 328.
%
\bibitem{ref:Xu-Grei}
Z.~Xu and C.~Greiner, Phys.~Rev.~{\bf D62} (2000) 036012.
%
\bibitem{ref:Risch} 
D.~H.~Rischke,~Phys.~Rev.~{\bf C58} (1998) 2331.
%
\bibitem{ref:KMT2}
T.~Koide, M.~Maruyama and F.~Takagi, Prog.~Theor.~Phys.~{\bf 107} (2002) 1001.
%
\bibitem{ref:Naka}
S.~Nakajima, Prog.~Theor.~Phys.~{\bf 20} (1958) 948.
%
\bibitem{ref:Zwanzig1}
R.~Zwanzig, J.~Chem.~Phys.~{\bf 33} (1960) 1338.
%
\bibitem{ref:Mori}
H.~Mori, 
Prog.~Theor.~Phys.~{\bf 33} (1965) 423.
%
\bibitem{ref:SH1}
N.~Hashitsume, F.~Shibata and M.~Shing\=u, 
J.~Stat.~Phys.~{\bf 17} (1977) 155.
%
\bibitem{ref:SH2}
F.~Shibata, Y.~Takahashi and N.~Hashitsume, 
J.~Stat.~Phys.~{\bf 17} (1977) 171.
%
\bibitem{ref:SH3}
F.~Shibata and T.~Arimitsu, J.~Phys.~Soc.~Jpn~{\bf 49} (1980) 891.
%
\bibitem{ref:SH4}
C.~Uchiyama and F.~Shibata, Phys.~Rev.~{\bf E60} (1999) 2636.
%
\bibitem{ref:KMT1}
T.~Koide, M.~Maruyama and F.~Takagi, Prog.~Theor.~Phys.~{\bf 101} (1999) 373.
%
\bibitem{ref:KM1}
T.~Koide and M.~Maruyama, 
Prog.~Theor.~Phys.~{\bf 104} (2000) 575.
%
\bibitem{ref:Koide}
T.~Koide, Prog.~Theor.~Phys.~{\bf 107} (2002) 525.
%
\bibitem{ref:KM2}
T.~Koide and M.~Maruyama, nucl-th/0308025.
%
\bibitem{ref:Rau1}
J.~Rau,~Phys.~Rev.~{\bf 50} (1994) 6911.
%
\bibitem{ref:Rau2}
J.~M\"uller and J.~Rau,~Phys.~Lett. {\bf B386} (1996) 274.
%
\bibitem{ref:Rau3}
P.~Neu and J.~Rau,~Phys.~Rev.~{\bf E55} (1997) 2195.
%
\bibitem{ref:Ayik}
S.~Ayik, Phys.~Rev.~Lett. {\bf 56} (1986) 38.
%
\bibitem{ref:Ana}
C.~Anastopoulos,~Phy.~Rev.~{\bf D56} (1997) 1009.
%
\bibitem{ref:Howard}
The structure of the memory function has been 
studied in the recurrence relations approach. 
See \cite{ref:review2}.
%
\bibitem{ref:review} 
J.~Rau and B.~M\"uller, Phys.~Rep.~{\bf 272} (1996) 1.
%
\bibitem{ref:review2}
U.~Balucanai, M. Howard Lee and V.~Tognetti, Phys.~Rep.~{\bf 373} (2003) 409.
%
\bibitem{ref:vanHove}
L.~van Hove, Phys.~Rev.~{\bf 95} (1954) 1374.
%
\bibitem{ref:HK-PR}
S.~P.~Klevansky, Rev.~Mod.~Phys.~{\bf 64} (1992) 649.
%
\bibitem{ref:HK-PR2}
T.~Hatsuda and T.~Kunihiro, Phys.~Rep.~{\bf 247} (1994) 222.
%
\bibitem{ref:footnote1}
Above the critical temperature, 
there is no difference between $\sigma$ and $\delta \sigma$.
However, it is more convenient to use $\delta \sigma$ 
instead of $\sigma$ in the symmetry broken phase.
%
\bibitem{ref:Sawada}
J.~Okada, I. Sawada and Y.~Kuroda, 
J.~Phys.~Soc.~Jpn~{\bf 64} (1995) 4092.
%
\bibitem{ref:Kubo-book}
R.~Kubo, M.~Toda and N.~Hashitsume, 
{\it Statistical Physics II} (Springer-Verlag, Berlin, 1983).
%
\bibitem{ref:Fick}
E.~Fick and G.~Sauermann,
{\it The Quantum Statistics of Dynamic Process} (Springer-Verlag, Berlin, 1983).
%
\bibitem{ref:KKKN3}
See, for example, 
M.~Kitazawa, T.~Koide, T.~Kunihiro and Y.~Nemoto,
Prog.~Theor.~Phys.~{\bf 108} (2002) 929;
Prog.~Theor.~Phys.~{\bf 110} (2003) 185 (addenda).
%
\bibitem{ref:Thou}
D.~J.~Thouless, Ann~Phys.~(N.Y.)~{\bf 10} (1960) 553.
%
\bibitem{ref:HK}
T.~Hatsuda and T.~Kunihiro, Phys.~Rev.~Lett.~{\bf 55} (1985) 158.
%
\bibitem{ref:KKKN}
M.~Kitazawa, T.~Koide, T.~Kunihiro and Y.~Nemoto,
Phys.~Rev.~{\bf D65} (2002) 091504(R).
%
\bibitem{ref:Fodor}
Z.~Fodor and S.~D.~Katz, 
Phys. Lett. {\bf B534} (2002) 87.
%
\bibitem{ref:Fodor2}
Z.~Fodor and S.~D.~Katz, 
JHEP {\bf 0203} (2002) 014.
%
\bibitem{ref:Karsch}
C.~R.~Allton, S.~Ejiri, S.~J.~Hands, O.~Kaczmarek, F.~Karsch, E.~Laermann, 
Ch.~Schmidt and L.~Scorzato,
Phys.~Rev.~{\bf D66} (2002) 074507.
%
\bibitem{ref:de}
P.~de Forcrand and O.~Philipsen, Nucl.~Phys.~{\bf B642} (2002) 290.
%
\bibitem{ref:Elia}
M.~D'Elia and M.~P.~Lombard, Phys.~Rev.~{\bf D67} (2003) 014505.
%
\bibitem{ref:de2}
P.~de Forcrand and O.~Philipsen, Nucl.~Phys.~{\bf B673} (2003) 170.
%
\bibitem{ref:Fodor3}
Z.~Fodor and S.~D.~Katz, hep-lat/0402006.
%
\bibitem{ref:glass}
W.~G\"otze, 
in {\it Liquids, Freezing, and the Glass Transition}, editted by 
D.~Levesque, J.~P.~Hansen and J.~Zinn-Justin (North Holland, Amsterdam, 1990).
%
\bibitem{ref:TFD}
H.~Umezawa, H.~Matsumoto and M.~Tachiki, 
{\it Thermo Field Dynamics and Condensed States} (North-Holland, Amsterdam, 1982).
%
\bibitem{ref:Zwanzig}
R.~Zwanzig, J.~Stat.~Phys.~{\bf 9} (1973) 215.
%
\end{thebibliography}
\end{document}